\documentclass[pdftex,twocolumn,epjc3]{svjour3}          

\RequirePackage[T1]{fontenc}

\smartqed  

\RequirePackage{graphicx}
\RequirePackage{mathptmx}      
\RequirePackage{flushend}
\RequirePackage[numbers,sort&compress]{natbib}
\RequirePackage[colorlinks,citecolor=blue,urlcolor=blue,linkcolor=blue]{hyperref}

\usepackage{amsmath}
\usepackage{braket}
\usepackage{siunitx}
\usepackage{csquotes}
\usepackage{bm}
\usepackage{xifthen}
\usepackage{microtype}

\DeclareSIUnit{\noop}{\kern 0pt}

\newcommand{\tr}{\ensuremath{\operatorname{Tr}}}
\newcommand*\dif{\mathop{}\!\mathrm{d}}
\newcommand{\imag}{\ensuremath{\text{i}}}
\newcommand{\T}{\ensuremath{\operatorname{T}}}
\newcommand{\LQCD}{\ensuremath{\Lambda_\text{QCD}}}
\newcommand{\Order}{\ensuremath{\mathcal{O}}}

\newcommand{\mpi}{\ensuremath{m_\pi}}
\newcommand{\etaetap}{\ensuremath{{\eta / \eta'}}}
\newcommand{\Q}{\mathcal{Q}}

\newcommand{\Pipion}[1][]{%
	\ifthenelse{\equal{#1}{}}%
	{\ensuremath{\bar{\Pi}_1^{(3)}}}%
	{\ensuremath{\bar{\Pi}_1^{(3),\ #1}}}%
}
\newcommand{\Pieta}[1][]{%
	\ifthenelse{\equal{#1}{}}%
	{\ensuremath{\bar{\Pi}_1^{(8/0)-\eta}}}%
	{\ensuremath{\bar{\Pi}_1^{(8/0)-\eta,\ #1}}}%
}
\newcommand{\Pietap}[1][]{%
	\ifthenelse{\equal{#1}{}}%
	{\ensuremath{\bar{\Pi}_1^{(8/0)-\eta'}}}%
	{\ensuremath{\bar{\Pi}_1^{(8/0)-\eta',\ #1}}}%
}
\newcommand{\PiPS}[1][]{%
	\ifthenelse{\equal{#1}{}}%
	{\ensuremath{\bar{\Pi}_1^{(a)}}}%
	{\ensuremath{\bar{\Pi}_1^{(a),\ #1}}}%
}
\newcommand{\Pietaetap}[1][]{%
	\ifthenelse{\equal{#1}{}}%
	{\ensuremath{\bar{\Pi}_1^{(8/0)-\etaetap}}}%
	{\ensuremath{\bar{\Pi}_1^{(8/0)-\etaetap,\ #1}}}%
}

\newcommand{\aPion}[1][]{%
	\ifthenelse{\equal{#1}{}}%
	{\ensuremath{a_\mu^{(3)}}}%
	{\ensuremath{a_\mu^{(3),\ #1}}}%
}

\newcommand{\aEta}[1][]{%
	\ifthenelse{\equal{#1}{}}%
	{\ensuremath{a_\mu^{(8/0)-\eta}}}%
	{\ensuremath{a_\mu^{(8/0)-\eta,\ #1}}}%
}

\newcommand{\aEtap}[1][]{%
	\ifthenelse{\equal{#1}{}}%
	{\ensuremath{a_\mu^{(8/0)-\eta'}}}%
	{\ensuremath{a_\mu^{(8/0)-\eta',\ #1}}}%
}

\newcommand{\aPS}[1][]{%
	\ifthenelse{\equal{#1}{}}%
	{\ensuremath{a_\mu^{(a)}}}%
	{\ensuremath{a_\mu^{(a),\ #1}}}%
}

\newcommand{\aEtaEtap}[1][]{%
	\ifthenelse{\equal{#1}{}}%
	{\ensuremath{a_\mu^{(8/0)-\etaetap}}}%
	{\ensuremath{a_\mu^{(8/0)-\etaetap,\ #1}}}%
}

\newcommand{\aLon}{\ensuremath{a_\mu^\text{long}}}

\renewcommand{\Re}{\ensuremath{\operatorname{Re}}}

\journalname{Eur. Phys. J. C}

\def\makeheadbox{{%
		\hbox to0pt{\vbox{\baselineskip=10dd\hrule\hbox
				to\hsize{\vrule\kern3pt\vbox{\kern3pt
						\hbox{This is a preprint of an article published in {\bfseries Eur.~Phys.~J.~C}.} \hbox{The final authenticated version is available online at: \href{https://doi.org/10.1140/epjc/s10052-020-08611-6}{DOI 10.1140/epjc/s10052-020-08611-6}.}
						\hbox{\vspace{-0.185cm}}
						\kern3pt}\hfil\kern3pt\vrule}\hrule}%
			\hss}}}

\begin{document}

\title{Effects of Longitudinal Short-Distance Constraints on the Hadronic Light-by-Light Contribution to the Muon $\mathbf{g-2}$}

\author{Jan L\"udtke\thanksref{e1,addr1}
        \and
        Massimiliano Procura\thanksref{e2,addr1}
}

\thankstext{e1}{e-mail: jan.luedtke@univie.ac.at}
\thankstext{e2}{e-mail: massimiliano.procura@univie.ac.at}

\institute{University of Vienna, Faculty of Physics, Boltzmanngasse 5, A-1090 Vienna, Austria\label{addr1}
}

\date{\vspace*{-2cm}}

\maketitle
\sloppy
\begin{abstract}
We present a model-independent method to estimate the effects of short-distance constraints (SDCs) on the hadronic light-by-light contribution to the muon anomalous magnetic moment $a_\mu^\text{HLbL}$. The relevant loop integral is evaluated using multi-parameter families of interpolation functions, which  satisfy by construction all constraints derived from general principles and smoothly connect the low-energy region with those where either two or all three independent photon virtualities become large. In agreement with other recent model-based analyses, we find that the SDCs and thus the infinite towers of heavy intermediate states that are responsible for saturating them have a rather small effect on $a_\mu^\text{HLbL}$. Taking as input the known ground-state pseudoscalar pole contributions, we obtain that the longitudinal SDCs increase $a_\mu^\text{HLbL}$ by  \num[separate-uncertainty]{9.1\pm5.0e-11}, where the isovector channel is responsible for \num[separate-uncertainty]{2.6\pm1.5e-11}. More precise estimates can be obtained with our method as soon as further accurate, model-independent information about  important low-energy contributions from hadronic states with masses up to \SIrange[range-units=single, range-phrase=--]{1}{2}{\GeV} become available.
\end{abstract}

\section{Introduction}

The persisting discrepancy between the Standard Model evaluation and the experimental determination~\cite{BNL} of the muon anomalous magnetic moment $a_\mu$ is one of the outstanding open problems in particle physics and is traditionally considered a harbinger of New Physics. Moreover, the forthcoming results from the Fermilab E989 experiment, which aim to improve the present accuracy by a factor of 4 to reach an uncertainty of about $16 \times 10^{-11}$ ({\it i.e.} 0.14 ppm)~\cite{FermilabTDR}, make it even more crucial and timely to further scrutinize and improve control over theory predictions.

Together with the hadronic vacuum polarization contribution, the hadronic light-by-light (HLbL) is the major source of theoretical uncertainty in the Standard Model~\cite{Prades:2009tw,Jegerlehner:2009ry,Jegerlehner:2008zza}. In the last years, significant efforts have been devoted to improve the determination of $a_\mu^{\text{HLbL}}$ and reduce model dependence by using analytic approaches based on dispersion relations~\cite{Hoferichter:2013ama,BTT1,Colangelo:2014pva,BTT2,Colangelo:2017qdm,BTT3,Hoferichter:2014vra,PionTFFshort,PionTFF,Pauk:2014rfa,Danilkin:2016hnh,Hagelstein:2017obr} as well as lattice QCD~\cite{Blum:2014oka,Blum:2015gfa,Blum:2016lnc,Blum:2017cer,Blum:2019ugy,Green:2015sra,Asmussen:2019act}. In particular, the dispersive framework for the HLbL tensor in Refs.~\cite{Hoferichter:2013ama,BTT1,Colangelo:2014pva,BTT2,Colangelo:2017qdm,BTT3} has enabled accurate data-driven determinations with controlled error estimates of the contributions from one- and two-pion intermediate states.

In this framework $a_\mu^{\text{HLbL}}$ is evaluated via a two-loop integral of dispersively reconstructed scalar functions against analytically known kernels. At sufficiently small space-like photon virtualities, contributions from low-mass states accessible to a dispersive treatment are enhanced. At higher virtualities such an enhancement does not occur leading to important effects from higher intermediate states, which are constrained by operator product expansions (OPEs) and perturbative QCD (pQCD).

More specifically, there are two relevant kinematic regimes concerning short-distance constraints (SDCs) on $a_\mu^{\text{HLbL}}$ for asymptotic values of (subsets of) the photon virtualities.
Since one of the photons corresponds to the static electromagnetic source in the definition of $g-2$, one asymptotic regime is realized when the remaining three space-like photon virtualities are comparable and much larger than $\LQCD^2$, and the other when two space-like photon virtualities are much larger than both the third and $\LQCD^2$. The latter SDC was first derived by Melnikov and Vainshtein (MV)~\cite{MV} using an OPE that leads to relations involving longitudinal and transversal amplitudes of the correlator of two vector and one axial current (VVA) in the chiral limit. The former SDC was also discussed in Ref.~\cite{MV} based on the quark-loop calculation at leading order in pQCD and its derivation was recently put on a firmer theoretical ground by means of an OPE in an external electromagnetic background field~\cite{Bijnens}.

Tree-level resonance exchanges cannot make $a_\mu^{\text{HLbL}}$ comply with all SDCs unless an infinite number of states is included. This is due to the fact that the transition form factors (TFFs) describing the resonance couplings to off-shell photons are subject themselves to asymptotic QCD constraints~\cite{Lepage:1980fj,Brodsky:1981rp,Hoferichter:2020lap},
which make the full HLbL four-point function decay too fast at high virtualities.\footnote{See {\it e.g.}\ Ref.~\cite{Bijnens:2003rc} where the analogous case of a three-point function is treated explicitly.} MV proposed a model to satisfy the longitudinal and transversal OPE SDCs through a modification of the pion pole contribution~\cite{MV, MVNew}, which affects also the low-energy region. Recently, alternative model-dependent solutions have been investigated to fulfill both OPE and pQCD SDCs by instead adding degrees of freedom to the ground-state pseudoscalars. In this context, Refs.~\cite{BernSDCShort,BernSDCLong} proposed the inclusion of infinite towers of excited pseudoscalar poles in large-$N_c$-inspired Regge models to satisfy longitudinal SDCs away from the chiral limit,\footnote{For other calculations based on large-$N_c$ arguments to satisfy long- and short-distance constraints on QCD correlators using finite or infinite sets of narrow resonances, see Refs. \cite{Peris:1998nj, Knecht:1998sp, Bijnens:2003rc, Golterman:2001nk, Knecht:2001xc, Golterman:2001pj, DAmbrosio:2019xph}.} while the effect of summing over axial-vector contributions in holographic QCD was the subject of   Refs.~\cite{HolographyVienna, HolographyItaly}.\footnote{See also Ref.~\cite{Masjuan:2020jsf} for a discussion of the role of axial-vector mesons in the saturation of the SDCs.} Through the explicit summation of intermediate states, these models provide specific interpolations between the low-energy region and the asymptotic regimes for the scalar functions that determine $a_\mu^{\text{HLbL}}$.

The goal of this paper is complementary to these studies. We introduce an approach based on more general interpolating scalar functions, independent of the physical mechanism that is ultimately responsible for their actual form outside the low-energy region. The multi-parameter families of functions studied here satisfy all constraints rigorously derived from general principles: unitarity, analyticity and crossing in the low-energy domain, OPE and pQCD constraints in the mixed and high-energy regions. Here we focus on longitudinal SDCs since these are tightly related to the pseudoscalar poles for which accurate low-energy input is available and 
their implementation does not involve any mixing of OPE constraints among different scalar functions~\cite{BernSDCLong}. Error estimates as well as the role played by the various parameters and assumptions, can be easily and transparently addressed in our approach and are investigated in detail in our numerical study. 

Crucial input for our analysis is provided by an accurate low-energy representation of the scalar functions. In the following we will mostly assume that this is given by the ground-state pseudoscalar poles. In this context, an important role is played by the lightest state ($\pi^0$), whose contribution is under firm theoretical control thanks to a dispersive evaluation~\cite{Hoferichter:2014vra,PionTFFshort,PionTFF}. Improved determinations of the effects of SDCs can be obtained in a straightforward way within our approach once similarly precise, model-independent information about further relevant intermediate states in the energy region up to \SIrange[range-units=single, range-phrase=--]{1}{2}{\GeV} become available. In order to illustrate this aspect and compare against a different way to estimate the contribution from SDCs, we have applied our method also to the case where the lightest axial-vector meson is included in the low-energy region using input from holographic QCD~\cite{HolographyVienna,HolographyItaly} and neglecting issues related to intrinsic model dependence.

The paper is structured as follows. In Sec.~\ref{Sec:SDC} we review the relevant constraints on HLbL and the assumptions made in their derivations. Sec.~\ref{Sec:HEInt} describes our interpolation between the OPE and pQCD asymptotic constraints while in Sec.~\ref{Sec:Int} we discuss its smooth connection with the low-energy region. In sec.~\ref{Sec:Num} we present our numerical analysis with particular emphasis on the error estimation. Conclusions are drawn in Sec.~\ref{Sec:Con}. \ref{App:Convergence} is devoted to the analysis of the convergence properties of our interpolants.
\section{Longitudinal short-distance constraints on HLbL}
\label{Sec:SDC}
\subsection{Master formula for \texorpdfstring{$a_\mu^\text{HLbL}$}{a mu HLbL} and pseudoscalar pole contributions}
\label{Sec:SDCBTT}
In order to set up the notation, we start by summarizing the relevant definitions and results from Refs.~\cite{BTT2,BTT3}. The HLbL contribution to $a_\mu$ is governed by the fourth-rank vacuum polarization tensor for fully off-shell photon-photon scattering in pure QCD, 
\begin{eqnarray}
\Pi^{\mu\nu\lambda\sigma}(q_1,q_2,q_3) & = & -\imag \int \dif^4 x\, \dif^4 y \,\dif^4 z \, e^{-\imag(q_1\cdot x + q_2 \cdot y + q_3 \cdot z)} \nonumber \\
&&\times \bra{0}\T\{j^\mu (x) j^\nu (y) j^\lambda (z) j^\sigma (0) \} \ket{0}
\label{Eq:HLbLTensor}
\end{eqnarray}
with  momenta assigned as $q_1+q_2+q_3 = q_4$. In this expression, the electromagnetic current for the light quark triplet is given by
\begin{eqnarray}
j^\mu (x) & = & \bar{\psi}(x) {\Q} \gamma^\mu \psi(x)\,, \nonumber \\
\psi & = & (u,d,s)^T\,, \nonumber \\
\Q & = & \frac{1}{3}\operatorname{diag}(2,-1,-1)\, .
\label{Eq:EM}
\end{eqnarray}

By generalizing the procedure introduced by Bardeen and Tung~\cite{BardeenTung} and Tarrach~\cite{Tarrach} in their studies of doubly-virtual Compton scattering, it is possible to derive a generating redundant \enquote{BTT} set of \num{54} Lorentz structures,
\begin{equation}
\Pi^{\mu\nu\lambda\sigma} = \sum_{i=1}^{54} T_i^{\mu\nu\lambda\sigma} \Pi_i\,,
\label{Eq:BTT}
\end{equation}
which is manifestly gauge invariant, closed with respect to crossing relations and such that the scalar functions $\Pi_i$ are free of kinematic singularities.

The HLbL contribution to $a_\mu$ can be derived from the tensor $\Pi^{\mu\nu\lambda\sigma}$ by using projection operator methods~\cite{Aldins:1970id,Barbieri:1974nc,Jegerlehner:2008zza} in the static limit $q_4 \to 0$. After performing a Wick rotation to Euclidean momenta, angular averages~\cite{Rosner:1967zz, Levine:1974xh} lead to the master formula~\cite{BTT3}
\begin{eqnarray}
a_\mu^\text{HLbL} & = & \frac{2\alpha^3}{3\pi^2}\int_{0}^{\infty} \dif Q_1 \int_{0}^{\infty} \dif Q_2 \int_{-1}^{1} \dif \tau \sqrt{1-\tau^2}\,Q_1^3\, Q_2^3\,\nonumber\\
&& \times \sum_{i=1}^{12} T_i(Q_1,Q_2,\tau) \,\bar{\Pi}_i(Q_1,Q_2,\tau)\,,
\label{Eq:MasterFQ}
\end{eqnarray}
where $Q_{1,2}$ denote the magnitudes of the Euclidean loop four-momenta, $Q_{1,2} = \sqrt{-q_{1,2}^2}$, and $\tau$ is the cosine of the angle between these vectors. The scalar functions $\bar{\Pi}_i$ are linear combinations of the previous $\Pi_i$ for $q_4 \to 0$. The analytic expressions of the integration kernels $T_i$ are given in Ref.~\cite{BTT3}.

Parameterizing the three-dimensional integration domain by the coordinates \cite{EichmannParametrization}
\begin{equation}
\Sigma \in [0,\infty)\,,\quad r \in [0,1]\,,\quad \phi \in [0,2\pi)\, ,
\end{equation}
which are related to the non-vanishing photon virtualities by
\begin{eqnarray}
Q_1^2 & = & \frac{\Sigma}{3}\left(1 - \frac{r}{2} \cos\phi - \frac{r}{2} \sqrt{3}\sin \phi\right)\,, \nonumber \\
Q_2^2 & = & \frac{\Sigma}{3}\left(1 - \frac{r}{2} \cos\phi + \frac{r}{2} \sqrt{3}\sin \phi\right)\,, \nonumber \\
Q_3^2 & = & Q_1^2 + 2 Q_1 Q_2 \tau + Q_2^2 = \frac{\Sigma}{3}\left(1 + r \cos\phi\right)\,,
\end{eqnarray}
will prove very useful in the following discussion about asymptotic constraints on HLbL. The master formula in Eq.~(\ref{Eq:MasterFQ}) then takes the form
\begin{eqnarray}
a_\mu^\text{HLbL} & = & \frac{\alpha^3}{432\pi^2}\int_{0}^{\infty} \dif \Sigma \,\Sigma^3 \int_{0}^{1} \dif r\, r\sqrt{1-r^2} \int_{0}^{2\pi} \dif \phi \nonumber\\
&& \times\sum_{i=1}^{12} T_i(\Sigma,r,\phi) \bar{\Pi}_i(\Sigma,r,\phi)\, .
\end{eqnarray}

In terms of the $Q_i^2$ coordinates, the integration domain amounts to a cone with tip at the origin. In terms of $(\Sigma, r, \phi)$, a given point in this cone is specified by the distance $\Sigma$ to the tip of the point's projection on the symmetry axis ($\Sigma = Q_1^2+Q_2^2+Q_3^2$), and by the polar coordinates $r$ and $\phi$ on the plane containing the point and orthogonal to the symmetry axis, normalized such that $r = 1$ corresponds to the surface of the cone.

In the master formula, a special role is played by the scalar functions $\bar{\Pi}_{1,2}$, which fulfill
\begin{equation}
\bar{\Pi}_2 = \mathcal{C}_{2,3}[\bar{\Pi}_1] \text{ and}\quad \mathcal{C}_{1,2}[\bar{\Pi}_1] = \bar{\Pi}_1\, ,
\label{Eq:crossing}
\end{equation}
where the crossing operator $\mathcal{C}_{i,j}$ exchanges momenta and Lorentz indices of the photons $i$ and $j$. These functions are the only ones describing the effects of pseudoscalar tree-level exchanges. For small values of $\Sigma$, the pion pole dominates yielding the largest contribution to $a_\mu^{\text{HLbL}}$ and also $\eta$/$\eta'$ poles yield sizable effects. Furthermore, distinctively, the OPE SDCs on $\bar{\Pi}_{1,2}$ do not involve other scalar functions~\cite{BernSDCLong}.

The functional form of $\bar{\Pi}_{1,2}$ in specific kinematic regimes is constrained according to analytic QCD results, which we will fully exploit to estimate the impact of (longitudinal) SDCs on
\begin{eqnarray}
\aLon & \equiv & \frac{\alpha^3}{432\pi^2}\int_{0}^{\infty} \dif \Sigma  \int_{0}^{1} \dif r   \int_{0}^{2\pi} \dif \phi \,\Sigma^3\, r\sqrt{1-r^2}
\nonumber \\
&& \times \left[T_1(\Sigma, r, \phi) + T_2\left(\Sigma, r, \phi + \frac{2\pi}{3}\right)\right]\bar{\Pi}_1(\Sigma, r, \phi) \,,
\label{Eq:MasterFPi1}
\end{eqnarray}
where the shift in the variable $\phi$ in $T_2$ corresponds to the crossing operation on $\bar{\Pi}_1$. Thus for our analysis, we only need to study one BTT scalar function in the $g-2$ kinematics.

For the purpose of later discussion, we stress here that a pole term in $\bar{\Pi}_1$ due to a single-particle intermediate state of mass $M$ yielding the denominator $Q_3^2 + M^2$ leads to a hierarchy among contributions in the space-like momentum region relevant for $a_\mu^\text{HLbL}$. For small values of $Q_3^2$ larger masses get suppressed, while for $Q_3^2$ comparable to the squared mass of the heavier state or larger, no suppression is expected.\footnote{This argument obviously also holds if the denominator of the heavier state gets replaced by $M^2$ as for the axial-meson longitudinal contribution in Eq.~(\ref{Eq:Axial}) below (see also Ref.~\cite{Masjuan:2020jsf}).} This effect is of course modified by the numerator in $\bar{\Pi}_1$, which encodes information on the strength of the coupling to two (off-shell) photons, but it still helps us identify which states \emph{can} be relevant at specific energy scales and which not, independent of the values of $Q_{1,2}^2$.

The lightest state contributing to $\bar{\Pi}_1$ is $\pi^0$. The unitarity relation for a single pseudoscalar intermediate state yields
\begin{equation}
\bar{\Pi}_1^\text{PS-pole} = -\frac{F_{\text{PS} \gamma^* \gamma^*}(-Q_1^2, -Q_2^2) F_{\text{PS} \gamma^* \gamma^*}(-Q_3^2, 0)}{Q_3^2 + m_\text{PS}^2}\, ,
\label{Eq:Pi1PSpole}
\end{equation}
where the numerator is given by the product of a doubly-virtual and a singly-virtual transition form factor (TFFs) for an on-shell pseudoscalar meson (PS), which is defined by the matrix element
\begin{eqnarray}
&&\imag \int \dif^4 x\, e^{\,\imag q_1\cdot x} \bra{0} \T\{j_\mu(x) j_\nu(0)\}\ket{\text{PS}(q_1+q_2)} \nonumber\\
&&\quad= \epsilon_{\mu\nu\alpha\beta} q_1^\alpha q_2^\beta \,F_{\text{PS}\gamma^*\gamma^*}(q_1^2,q_2^2)
\label{Eq:TFF}
\end{eqnarray}
with $\epsilon^{0 1 2 3}=+1$. If we set $\bar{\Pi}_1 = \bar{\Pi}_1^\text{PS-pole}$, then $\aLon$ amounts to the pseudoscalar pole contribution $a_\mu^{\text{PS-pole}}$. In the $\pi^0$ case, this has been evaluated within a few percent accuracy via a data-driven dispersive approach~\cite{Hoferichter:2014vra,PionTFFshort,PionTFF},
\begin{equation}
a_{\mu,\text{disp}}^{\pi^0\text{-pole}} = 62.6^{+3.0}_{-2.5}\times 10^{-11}\, .
\label{Eq:aPionPole}
\end{equation}
This result agrees with other recent determinations based on lattice QCD~\cite{Gerardin:2019vio}, Canterbury approximants~\cite{Canterbury}, Dyson-Schwinger equations~\cite{DSE,Raya:2019dnh} and AdS/QCD models~\cite{Leutgeb:2019zpq}. While a dispersive analysis of the doubly-virtual $\eta$/$\eta'$ TFFs has not been completed yet,\footnote{The dispersive formalism for the singly-virtual $\eta$/$\eta'$ TFF has been established~\cite{Hanhart:2013vba} and progress has been made towards the determination of the doubly-virtual isovector contribution~\cite{Xiao:2015uva,Kubis:2018bej}.} the method of Canterbury approximants in Ref.~\cite{Canterbury} provides data-driven determinations 
and associated uncertainties also for the $\eta$/$\eta '$ TFFs.  In our numerical analysis of SDCs, we have used as input the dispersive $\pi^0$ TFF from Refs.~\cite{PionTFFshort,PionTFF} and compared our final results against those with form factors from Canterbury and Dyson-Schwinger approaches, while for $\eta/\eta'$ we have used the TFFs in Ref.~\cite{Canterbury} and compared against Ref.~\cite{DSE}.

\begin{figure}
	\centering
	\includegraphics{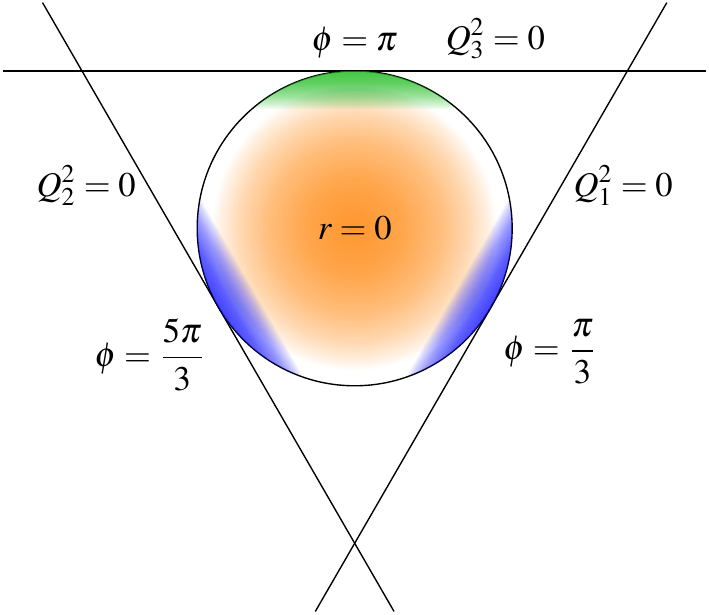}
	
	\caption{The circle represents the boundary of the $g-2$ integration domain for a fixed value of $\Sigma$. The angles $\phi = \pi/3$, $\phi = \pi$ and $\phi=5 \pi/3$ correspond to $Q_2^2= Q_3^2$, $Q_1^2 = Q_2^2$ and $Q_1^2=Q_3^2$, respectively. The colored regions denote where SDCs on $\bar{\Pi}_1$ hold at large $\Sigma$. The blue domains yield contributions to $\bar{\Pi}_1$ from the OPE expansion of the VVA correlator that are sub-leading compared to the green one, while the orange region corresponds to the pQCD constraint.}
	\label{Fig:patches}
\end{figure}

The asymptotic constraints on $a_\mu^\text{HLbL}$~\cite{MV,Bijnens}  (see also Refs.~\cite{Knecht:2020xyr,Masjuan:2020jsf}) that we are going to discuss in the next sections have been translated into the BTT framework in Refs.~\cite{Bijnens,BernSDCShort,BernSDCLong}.
In this context, there are two distinct relevant kinematic regimes. The first (asymmetric) one is realized when one of the photon virtualities is much smaller than the other two, which are large and comparable, {\it e.g.}\ $Q_1^2 \sim Q_2^2 \gg Q_3^2$. The second (symmetric) limit occurs when all the Euclidean non-vanishing photon virtualities are large and comparable in size ($Q_1^2 \sim Q_2^2 \sim Q_3^2 \gg \LQCD^2 $). Both asymptotic limits correspond to $\Sigma \to \infty$ but for different values of $r$ and $\phi$: the asymmetric limit $Q_1^2 \sim Q_2^2 \gg Q_3^2$ corresponds to $r=1$ and $\phi = \pi$ while the symmetric configuration occurs in a neighborhood of $r=0$ (see Fig.~\ref{Fig:patches}). 

In the following we will review the relevant constraints on $\bar{\Pi}_1$ at large $\Sigma$ and describe in detail our method to provide general families of interpolants for $\bar{\Pi}_1(\Sigma, r, \phi)$ between low- and high-energy regions in the $g-2$ integral.

\subsection{The asymmetric asymptotic region: OPE constraints}
\label{Sec:OPE}
For large Euclidean values of $\hat{q} \equiv (q_1 - q_2)/2$, one can expand the time-ordered product of two electromagnetic currents, which defines the tensor
\begin{equation}
\Pi^{\mu\nu}(q_1, q_2) = \imag \int \dif^4 x\, \dif^4 y \,e^{-\imag(q_1\cdot x + q_2 \cdot y)}\T\{j^\mu (x) j^\nu (y)\}\,,
\label{Eq:Pimunu}
\end{equation}
into a series of local operators. At leading order in $\alpha_s$, by matching single-quark matrix elements and omitting the unit operator, which does not contribute to the connected HLbL tensor in Eq.~(\ref{Eq:HLbLTensor}), one obtains~\cite{Bjorken:1966jh}
\begin{eqnarray}
&&\Pi^{\mu\nu}(q_1, q_2)\nonumber\\
&&\quad = \int \dif^4 z\, e^{-\imag(q_1 + q_2) \cdot z} \left(-\frac{2\imag}{\hat{q}^2}\epsilon^{\mu\nu\alpha\beta} \hat{q}_\alpha j_{5\beta}(z)\right) + \dots\, ,
\label{Eq:OPE}
\end{eqnarray}
where the axial current $j_5^\mu$ is defined by $j_5^\mu(x)=\bar{\psi}(x)\Q^2 \gamma^\mu \gamma_5 \psi(x)$ with charge matrix given in Eq.~(\ref{Eq:EM}). The ellipsis denotes sub-leading terms suppressed by powers of $\{|q_1+q_2|/|\hat{q}|,\,\allowbreak\Lambda_{\rm QCD}/|\hat{q}|\}$. This result implies that, at leading order in the OPE and at leading order in $\alpha_s$, the HLbL tensor can be expressed in terms of the correlator of two vector currents with an axial current,
\begin{eqnarray}
\Pi_{\mu\nu\lambda\sigma}(q_1,q_2,q_3) & = & \frac{2 \imag}{\hat{q}^2} \epsilon_{\mu\nu\alpha\beta} \hat{q}^\alpha \int \dif^4 x\dif^4 y\, e^{-\imag q_3\cdot x} e^{\imag q_4\cdot y} \nonumber \\
&& \times\bra{0} \T \{j_\lambda(x) j_\sigma(y) j_5^\beta(0)\}\ket{0} + \dots \nonumber \\
\end{eqnarray}
for $Q_1^2 \sim Q_2^2 \gg \{Q_3^2,Q_4^2,\LQCD^2\}$. This three-point function, which also appears in the calculation of fermion loop electroweak contributions to $a_\mu$~\cite{Knecht:2002hr,VVAold}, can be decomposed into Lorentz structures that are longitudinal and transversal with respect to the Lorentz index of the axial current (see {\it e.g.}\ Ref.~\cite{VVA}). The corresponding longitudinal scalar function determines the asymptotic behavior of $\bar{\Pi}_1$ in the asymmetric region and is fixed by the axial Adler--Bell--Jackiw anomaly up to chiral corrections and the gluon anomaly. Neglecting these effects, which will be discussed below, this translates into the following constraint~\cite{BernSDCLong} for the singlet and octet flavor components of $\bar{\Pi}_1(Q_1^2, Q_2^2, Q_3^2)$, defined by the decomposition of the axial current,
\begin{equation}
\bar{\Pi}_1^{(a), {\rm OPE}}(Q^2, Q^2, Q_3^2) = -\frac{2 N_c \,C_a^2}{\pi^2 Q^2 Q_3^2}\quad {\rm for} \quad  a=\{3,8,0\}\,,
\label{Eq:OPEConstraint}
\end{equation}
where $C_a = \tr(\Q^2 \lambda_a)/2$ in terms of the charge matrix $\Q$  and Gell-Mann matrices $\lambda_a$. In particular,
\begin{equation}
C_3 = \frac{1}{6}\, ,\quad C_8 = \frac{1}{6\sqrt{3}}\, ,\quad C_0 = \frac{2}{3\sqrt{6}}\,.
\label{Eq:Ca}
\end{equation}
Since Eq.~(\ref{Eq:OPEConstraint}) relies on a perturbative calculation of the VVA correlator, it holds in the kinematic limit $Q_1^2 \sim Q_2^2 \equiv Q^2 \gg Q_3^2 \gg \LQCD^2$. For the non-singlet components ($a=3,8$), since perturbative~\cite{Adler:1969er} as well as non-perturbative~\cite{tHooft:1979rat, Witten:1983tw} corrections are absent in the chiral limit, the hierarchy between $\LQCD^2$ and $Q_3^2$ can be dropped. In contrast, the singlet channel ($a=0$) is affected by the gluon anomaly, even in the chiral limit. This does not modify Eq.~(\ref{Eq:OPEConstraint}) for $Q_3^2 \gg \LQCD^2$~\cite{BernSDCLong}, but the extrapolation to small $Q_3^2$ is only valid if in addition the large-$N_c$ limit is considered, where the anomaly vanishes. Furthermore, in the crossed kinematics ($Q_2^2 \sim Q_3^2 \gg Q_1^2$ and $Q_1^2 \sim Q_3^2 \gg Q_2^2$), the leading-order OPE contributions to $\bar{\Pi}_1$ vanish.

Let us now compare $\PiPS[\text{OPE}]$ against the pseudoscalar pole contributions, focusing on the pion pole first. In the chiral limit and using the fact that $\lim\limits_{Q^2\to\infty} Q^2 F_{\pi\gamma^*\gamma^*}(-Q^2, -Q^2) = 4 C_3 F_\pi$ at leading order in $\alpha_s$~\cite{Novikov:1983jt,Manohar:1990hu}, one finds
\begin{minipage}{\columnwidth}
\begin{equation}
\lim\limits_{Q^2 \to \infty} Q^2 \bar{\Pi}_1^\text{$\pi^0$-pole}(Q^2, Q^2, Q_3^2) = -4 C_3 F_\pi \frac{F_{\pi\gamma^*\gamma^*}(-Q_3^2, 0)}{Q_3^2}\, .
\end{equation}
\end{minipage}
This expression has a pole at $Q_3^2 = 0$ since $F_{\pi\gamma^*\gamma^*}(0, 0) = 3 C_3 / (2\pi^2 F_\pi)$. The location of this pole as well as its residue agree with $\Pipion[\text{OPE}]$ in Eq.~(\ref{Eq:OPEConstraint}) (cf.~Ref.~\cite{MV}), which is consistent with the pion being the only massless isovector state in the chiral limit.

For finite quark masses, the pole in $\bar{\Pi}_1^\text{$\pi^0$-pole}$ is shifted from $Q_3^2 = 0$ to $Q_3^2 = -\mpi^2$, which lies outside the integration domain for $a_\mu^{\text{HLbL}}$. The closest point in the integration region for fixed asymptotic $\Sigma$ (see Fig.~\ref{Fig:patches}) is at $Q_3^2 = 0$, where
\begin{eqnarray}
\lim\limits_{Q^2 \to \infty} Q^2 \bar{\Pi}_1^\text{$\pi^0$-pole}(Q^2, Q^2, 0) & = & -4 C_3 F_\pi \frac{F_{\pi\gamma^*\gamma^*}(0, 0)}{\mpi^2} \nonumber \\ & = & - \frac{6 C_3^2}{\pi^2 \mpi^2}\, .
\label{Eq:PiPoleVanQ3}
\end{eqnarray}
Here few percent chiral corrections to $F_{\pi\gamma^*\gamma^*}(0, 0)$~\cite{Moussallam:1994xp, Goity:2002nn, Ananthanarayan:2002kj, Ioffe:2007eg, Kampf:2009tk} have been neglected. This is still close to the actual pole, which leads to the enhancement by $\mpi^{-2}$. Since no other contribution receives the same enhancement, the last expression is expected to provide an excellent approximation to the true $\Pipion$ in the specified limit.\footnote{At variance with the MV model of Ref.~\cite{MV}, we do not neglect the momentum dependence of the singly-virtual TFF and we allow for the contribution from other states besides the pion at finite $Q_3^2$.} We observe that the OPE result, which is derived in the chiral limit, reproduces Eq.~(\ref{Eq:PiPoleVanQ3}) if the pole position is shifted by the pion mass as dictated by the pion pole contribution
\begin{equation}
\lim\limits_{Q^2\to\infty} Q^2 \Pipion(Q^2, Q^2, Q_3^2) = -\frac{6 C_3^2}{\pi^2 (Q_3^2 + \mpi^2)}\, .
\label{Eq:OPEPionMass}
\end{equation}
This is also consistent with the OPE result in Eq.~(\ref{Eq:OPEConstraint}) for $Q^2 \gg Q_3^2 \gg \LQCD^2$, where chiral corrections are sub-leading. Thus, Eq.~(\ref{Eq:OPEPionMass}) is exact for $Q_3^2 \gg \LQCD^2$, relies on the assumption of pion dominance at $Q_3^2 \ll \LQCD^2$ and has the correct chiral limit Eq.~(\ref{Eq:OPEConstraint}) for all $Q_3^2$.  We extend it to the $\eta$/$\eta'$ channels and write
\begin{equation}
\lim\limits_{Q^2\to\infty} Q^2 \PiPS(Q^2, Q^2, Q_3^2) = -\frac{6 C_\text{PS}^2}{\pi^2 (Q_3^2 + m_\text{PS}^2)}\, .
\label{Eq:OPEConstraintMasses}
\end{equation}
Here $C_\pi = C_3$ but $C_\etaetap$ cannot be directly identified with $C_{0/8}$ due to $\eta$-$\eta'$-mixing. In analogy to the pion channel, we assume that ground-state single-pseudoscalar exchanges dominate $\Pietaetap(Q^2, Q^2, 0)$, despite the fact that the $\eta$/$\eta'$ poles are further away from $Q_3^2 = 0$. This assumption implies that $C_\etaetap$ can be read off from the pole contributions
\begin{equation}
\lim\limits_{Q^2\to\infty} Q^2\bar{\Pi}_1^{\text{$\etaetap$-pole}}(Q^2, Q^2, 0) = -\frac{6 C_{\etaetap}^2}{\pi^2 m_{\etaetap}^2}\,.
\label{Eq:defCetaetap}
\end{equation}

One can show that in the chiral limit and neglecting the gluon anomaly~\cite{BernSDCLong}
\begin{eqnarray}
&&\lim\limits_{Q_3^2\to 0} \, \lim\limits_{Q^2\to\infty} Q^2 Q_3^2 \nonumber \\ &&\times\left(\bar{\Pi}_1^{\eta\text{-pole}}(Q^2, Q^2, Q_3^2) + \bar{\Pi}_1^{\eta'\text{-pole}}(Q^2, Q^2, Q_3^2)\right) \nonumber \\ &&\quad= -\frac{6 (C_8^2 + C_0^2)}{\pi^2}\,.
\label{Eq:mixingOPE}
\end{eqnarray}
At this point we note that, besides the $\alpha_s$ corrections to the TFFs and OPE coefficient discussed in Sec.~\ref{Sec:MVpert}, which affect all ground-state pseudoscalars in the same way, the gluon anomaly induces a running of the flavor singlet decay constant~\cite{Leutwyler:1997yr,Kaiser:1998ds,Kaiser:2000gs}. This running leads to an incomplete cancellation between the decay constants in the symmetric asymptotic and the real photon limits, which has a sizable impact due to the large scale separation~\cite{Agaev:2014wna,PabloPhD}. 

Since $\bar{\Pi}_1^{\text{$\etaetap$-pole}}$ can be expressed in terms of TFFs according to Eq.~(\ref{Eq:Pi1PSpole}), assuming that corrections due to non-vanishing meson masses are negligible both in the real photon limit and in the symmetric asymptotic limit of the $\etaetap$ TFFs, Eqs.~(\ref{Eq:OPEConstraintMasses}--\ref{Eq:mixingOPE}) together imply
\begin{equation}
C_\eta^2 + C_{\eta'}^2= C_8^2 + C_0^2
\label{Eq:CPhysFlavor}
\end{equation}
up to the above-mentioned anomaly-induced scale-dependence, which leads to a violation of this equality (cf.\ Sec.~\ref{Sec:NumIS}).

For $Q_3^2 \gg \LQCD^2$, the additional $Q_3^2$-suppression of the singly-virtual TFF leads to a mismatch between the pseudoscalar pole contributions and the OPE constraint. In Ref.~\cite{MV} MV proposed to solve this issue by setting the singly-virtual TFF equal to a constant. This prescription is not compatible with the dispersive definition of the pole contributions in the framework summarized in Sec.~\ref{Sec:SDCBTT}, according to which, instead, an infinite tower of heavier intermediate states is needed to saturate the constraint (see {\it e.g.}\ Ref.~\cite{BernSDCLong}). For this purpose, summations of series of contributions from excited pseudoscalars~\cite{BernSDCShort,BernSDCLong} and axials~\cite{HolographyVienna,HolographyItaly} have been recently performed in the context of hadronic models. In Secs.~\ref{Sec:NumResults} and~\ref{Sec:NumAxials}, we will compare the outcome of our analysis against these estimates of the effects of longitudinal SDCs.


\subsection{\texorpdfstring{$\alpha_s$}{alpha\_s} corrections to the OPE}
\label{Sec:MVpert}

The derivation of Eq.~(\ref{Eq:OPE}) has been performed at leading order in $\alpha_s$. Since no other operator of dimension 3 can appear in that OPE, $\alpha_s$ corrections only affect the OPE coefficient of the axial-vector current. In Refs.~\cite{Kodaira:1978sh, Kodaira:1979ib, Kodaira:1979pa}, this coefficient has been calculated to next-to-leading order (NLO). Including this contribution in Eq.~(\ref{Eq:OPE}) leads to
\begin{eqnarray}
\Pi^{\mu\nu}(q_1, q_2) & = & \int \dif^4 z \,e^{-\imag(q_1 + q_2) \cdot z} \nonumber \\
&&\times\bigg(-\frac{2\imag}{\hat{q}^2} \left(1 - \frac{\alpha_s}{\pi}\right) \epsilon^{\mu\nu\alpha\beta} \hat{q}_\alpha j_{5\beta}(z) \nonumber\\
&&\qquad+ \Order\left(\hat{q}^{-2}\right)\bigg)\,.
\label{Eq:OPENLO}
\end{eqnarray}
It follows that the NLO version of Eq.~(\ref{Eq:OPEConstraintMasses}) reads
\begin{equation}
\lim\limits_{Q^2 \to \infty} Q^2 \PiPS(Q^2, Q^2, Q_3^2) = -\frac{6 C_\text{PS}^2 }{\pi^2 (Q_3^2 + m_\text{PS}^2)}\left(1-\frac{\alpha_s}{\pi}\right)\, .
\label{Eq:OPEConstraintNLO}
\end{equation}

The two-current operator product not only enters the HLbL tensor, but also the pion TFF (see Eq.~(\ref{Eq:TFF})). Thus, any perturbative correction to the OPE Wilson coefficient automatically implies the same perturbative correction to the symmetric limit of the pion TFF and vice versa. In fact, the symmetric asymptotic pion TFF has been calculated to NLO in Refs.~\cite{Braaten,PionTFF},\footnote{In Ref.~\cite{Braaten} the hard scattering kernel has been computed to NLO. In the limit $Q_1^2 = Q_2^2$ this is independent of the momentum fraction carried by the interacting quark, which makes the result independent of the pion distribution amplitude.}
\begin{equation}
F_{\pi\gamma^*\gamma^*}(-Q^2,-Q^2) = \left(1 - \frac{\alpha_s}{\pi} + \Order\left(\alpha_s^2\right)\right)\frac{2 F_\pi}{3Q^2} + \Order\left(Q^{-4}\right)\,,
\end{equation}
which is consistent with Eq.~(\ref{Eq:OPENLO}). The fact that the $\alpha_s$ corrections agree between the HLbL tensor in the asymmetric asymptotic limit and the symmetric asymptotic pion TFF guarantees that the pion pole saturates $\Pipion$ at $Q_3^2 = 0$ in the chiral limit also beyond leading order in $\alpha_s$.

A comment on the renormalization scale dependence of the terms in Eq.(\ref{Eq:OPENLO}) is in order here. The non-singlet components of the axial current are conserved (up to quark mass corrections) and thus their anomalous dimensions vanish. This is not true for the singlet component due to the gluon anomaly~\cite{Espriu:1982bw, Bos:1992nd}, but we neglect this effect here because it starts at $\Order(\alpha_s^2)$. Therefore, since the perturbatively expanded dimensionless part $d$ of the Wilson coefficient is scale ($\mu$) independent,
\begin{equation}
d\left(-\frac{\hat{q}^2}{\mu^2}, \alpha_s(\mu^2)\right) = d(1, \alpha_s(-\hat{q}^2)) 
\end{equation}
and the terms $\alpha_s^n \ln^{n-1} (-\hat{q}^2/\mu^2)$ ($n\ge 1$) can be resummed using as input the $\beta$-function and the one-loop result with $\alpha_s$ evaluated at the scale $-\hat{q}^2$ (see also~\cite{Narison:1992fd}).


\subsection{The symmetric asymptotic limit: perturbative QCD constraints}
\label{Sec:pQCD}
In Ref.~\cite{Bijnens} it has been shown that the pQCD quark loop is the leading term of an OPE in the kinematic limit $Q_1^2 \sim Q_2^2 \sim Q_3^2 \gg \LQCD^2$, where the fourth (external) photon has vanishing momentum in $(g-2)$-kinematics. At leading order in this OPE and at leading order in $\alpha_s$~\cite{BernSDCLong}, 
\begin{eqnarray}
\bar{\Pi}_1^\text{pQCD} (q_1^2, q_2^2, q_3^2) & = & \frac{N_c \tr \Q^4}{16\pi^2} \int_{0}^{1} \dif x \int_{0}^{1-x} \dif y \, I_1(x,y) \nonumber \\
& = & \frac{1}{24\pi^2} \int_{0}^{1} \dif x \int_{0}^{1-x} \dif y\,  I_1(x,y)\,, \nonumber \allowdisplaybreaks\\
I_1(x,y) & = & -\frac{16 x(1-x-y)}{\Delta_{132}^2} \nonumber\\
&&- \frac{16 x y(1-2x)(1-2y)}{\Delta_{132} \Delta_{32}}\,, \nonumber \\
\Delta_{ijk} & = & m_q^2-xy q_i^2 -x(1-x-y)q_j^2 \nonumber \\
&&- y(1-x-y) q_k^2\,, \nonumber \\
\Delta_{ij} & = & m_q^2-x(1-x)q_i^2 - y(1-y) q_j^2\,.
\label{Eq:pQCDIntegral}
\end{eqnarray}

In the symmetric limit, neglecting terms that are suppressed by powers of $m_q^2/Q^2$,
\begin{eqnarray}
\bar{\Pi}_1^\text{pQCD} (Q^2, Q^2, Q^2) & = & \sum_{a={3,8,0}} \PiPS[\text{pQCD}](Q^2, Q^2, Q^2) \nonumber \\
& = & \sum_{a={3,8,0}} -  \frac{4 N_c C_a^2}{3 \pi^2 Q^4}\, ,
\label{Eq:pQCDConstraint}
\end{eqnarray}
where we have chosen to adopt the same flavor decomposition as for the asymmetric OPE case, Eq.~(\ref{Eq:OPEConstraint}). If higher-order perturbative corrections are small, the leading-order result above is expected to be a good approximation also away from the fully symmetric configuration as long as large logarithms of ratios of momenta are absent.

Since $\bar{\Pi}_1^\text{PS-pole}$ decays like $Q^{-6}$, (towers of) ha\-dro\-nic contributions beyond ground-state pseudoscalar poles have to be responsible for the behavior shown by Eq.~(\ref{Eq:pQCDConstraint}). Following the MV prescription in Ref.~\cite{MV}, the parametric dependence on $Q$ can be reproduced but with an incorrect coefficient.

In order to saturate the pQCD result in the isosinglet channels, we need coefficients $C_\etaetap^\text{pQCD}$ satisfying 
\begin{equation}
C_8^2 + C_0^2 = \left(C^\text{pQCD}_{\eta}\right)^2 + \left(C^\text{pQCD}_{{\eta'}}\right)^2\,.
\label{Eq:CPhysFlavorQCD}
\end{equation}
Since Eq.~(\ref{Eq:CPhysFlavor}) is violated, we define
\begin{equation}
\left(C_\etaetap^\text{pQCD}\right)^2 = (1 + \delta_0) C_\etaetap^2\, ,
\label{Eq:delta0}
\end{equation}
where the parameter $\delta_0$ is chosen such that Eq.~(\ref{Eq:CPhysFlavorQCD}) holds.
\section{Interpolating between asymptotic constraints}
\label{Sec:HEInt}

We approximate the true $\bar{\Pi}_1(\Sigma, r, \phi)$ following a two-step procedure. We first select functional forms that are valid for asymptotic $\Sigma$ and are compatible with the constraints discussed in the previous section. We then interpolate between this set of functions and various representations of $\bar{\Pi}_1$ at small $\Sigma$ determined by single-particle intermediate states. Here we work at leading order in $\alpha_s$. Perturbative corrections will be discussed in our numerical analysis in Sec.~\ref{Sec:Num}.

The relevant constraints on $\bar{\Pi}_1$ at large $\Sigma$ are given by Eq.~(\ref{Eq:OPEConstraintMasses}) for $Q_1^2 = Q_2^2 \gg Q_3^2$ and Eq.~(\ref{Eq:pQCDConstraint}) for $Q_1^2 = Q_2^2 = Q_3^2$. Both expressions as well as the vanishing result of the leading-order OPE contribution in the crossed kinematics are compatible with
\begin{eqnarray}
\PiPS[\text{asymp}'] & = & -\frac{4 N_c C_\text{PS}^2}{\pi^2(Q_3^2 + m_\text{PS}^2)(Q_1^2 + Q_2^2 + Q_3^2)} \nonumber \\
& = & -\frac{12 N_c C_\text{PS}^2}{\pi^2 \Sigma (3m_\text{PS}^2 + \Sigma + \Sigma r \cos\phi)}
\label{Eq:AsymPiPrime}
\end{eqnarray}
if $ C_\text{PS} = C_\text{PS}^\text{pQCD} $. Thus, Eq.~(\ref{Eq:AsymPiPrime}) interpolates between symmetric and asymmetric asymptotic limits. According to Sec.~\ref{Sec:pQCD}, $\delta_0$ parameterizes the anomaly corrections to the singlet VVA correlator and the resulting shift in $C_\text{PS}^\text{pQCD}$ with respect to $C_\text{PS}$. Since a term proportional to $\Sigma^{-2}$ and independent of $(r, \phi)$ does not change the leading behavior at $Q_3^2 = 0$ and thus does not spoil compatibility with the OPE constraint, we subtract $36 \delta_0 C_\etaetap^2/(\pi^2 \Sigma^2)$ from Eq.~(\ref{Eq:AsymPiPrime}) in the case of $\eta$/$\eta'$.

Obviously, the choice made in Eq.~(\ref{Eq:AsymPiPrime}) and the exact form of the singlet correction are not unique and we are free to add a generic function such that the interpolant still satisfies the constraints. In order to have a non-negligible effect at asymptotic values of $\Sigma$, this additional function should also scale as $\Sigma^{-2}$ and we demand it to be finite and analytic for all $r\le 1$.\footnote{$\bar{\Pi}_1$ cannot decay more slowly than $\Sigma^{-2}$ for any $(r,\phi)$ region in order for the $a_\mu$ integral in Eq.~(\ref{Eq:MasterFPi1}) to be finite.} Therefore it can be approximated by a Taylor series in $r\cos\phi$ and $r \sin\phi$ truncated after order $M$,
\begin{eqnarray}
\Pipion[\text{asymp}] & = & \Pipion[\text{asymp}'] +\frac{12 N_c C_\pi^2}{\pi^2 \Sigma^2} \sum_{i=0}^{M} \sum_{j=0}^{M} \frac{1}{i!\ j!} \nonumber\\
&&\times a_{i,j} (r\cos\phi)^i (r\sin\phi)^j\,, \nonumber \\
\Pietaetap[\text{asymp}] & = & \Pietaetap[\text{asymp}'] - \frac{36 \delta_0 C_\etaetap^2}{\pi^2 \Sigma^2} \nonumber \\
&& + \frac{12 N_c C_\etaetap^2}{\pi^2 \Sigma^2} \sum_{i=0}^{M} \sum_{j=0}^{M} \frac{1}{i!\ j!} \nonumber \\
&&\times a_{i,j} (r\cos\phi)^i (r\sin\phi)^j\, ,
\label{Eq:AsymPi}
\end{eqnarray}
where
\begin{equation}
a_{0,0} = 0\,, \quad a_{i,2j+1} = 0
\label{Eq:Conaij}
\end{equation}
for integer $j$, due to the pQCD constraint and crossing symmetry. 

Up to now, we have applied the quark-loop result only at $r = 0$. However, the fact that Eq.~(\ref{Eq:pQCDIntegral}) holds also in a neighborhood of this point can be used to fix the coefficients $a_{i,j}$. To this end, we fitted Eq.~(\ref{Eq:AsymPi}) at fixed asymptotic $\Sigma$ with $M = 2$ to Eq.~(\ref{Eq:pQCDIntegral}) for $r<0.9$.\footnote{Since the maximal ratio of two squared momenta for $r=0.9$ is \num{14.7} and $\ln(14.7) \approx 2.7$, large logarithms do not occur in this region.} We chose a grid of equally separated points in this fitting region and minimized the sum of the relative squared differences between our interpolant and the leading-order quark-loop expression. The resulting 5 dimensionless fit parameters in the pion channel read
\begin{eqnarray}
a_{1,0} & = & \num{-0.170} \,,\quad a_{2,0} = \num{0.094} \,,\quad a_{0,2} = \num{-0.554}\,,\nonumber\\
\quad a_{1,2} & = & \num{-0.169} \,,\quad a_{2,2} = \num{-0.756}
\label{Eq:AsymPi_a_pion}
\end{eqnarray}
and are all at most $\Order(1)$, as expected since in Eq.~(\ref{Eq:AsymPi}) they parameterize relative corrections. This holds true also for the $\etaetap$ channels, where the numerical values are different. In Sec.~\ref{Sec:AsymUnc} we will discuss uncertainties due to the chosen fitting range, the number of parameters in the fit and $\alpha_s$ corrections to the asymptotic constraints.
\section{Interpolating between low and high energies}
\label{Sec:Int}
The next step is to smoothly connect our representation of $\bar{\Pi}_1$ for $\Sigma \gg \{\LQCD^2,M_\text{PS}^2,\dots\}$ given by Eq.~(\ref{Eq:AsymPi}) to an accurate low-energy description. We achieve this by adding suitable terms to $\PiPS[\text{asymp}]$ that are sub-leading at large $\Sigma$. For each choice of $r$ and $\phi$, the coefficients of these terms are then matched onto an input low-energy representation of $\bar{\Pi}_1$ at a suitably defined surface $\Sigma^\text{match}(r,\phi)$. In Sec.~\ref{Sec:Sigmamatch} we will discuss how $\Sigma^\text{match}$ is related to the mass scale at which intermediate states beyond the ones explicitly considered start to affect $\bar{\Pi}_1$.

\subsection{Interpolation functions and matching procedure}
For $\Sigma > \Sigma^\text{match}$ we consider the following two interpolation functions
\begin{eqnarray}
\PiPS[\text{int 1}](\Sigma, r, \phi) & = & \PiPS[\text{asymp}](\Sigma, r, \phi) \nonumber\\
&&\times \left(1 + \sum_{i = 1}^{N}b_i(r, \phi) \Sigma^{-i}\right)\,, \nonumber \\
\PiPS[\text{int 2}](\Sigma, r, \phi) & = & \PiPS[\text{asymp}](\Sigma, r, \phi) \nonumber\\
&&\times \left(1 + \sum_{i = 1}^{N}b_i(r, \phi) \Sigma^{-i}\right)^{-1}
\label{Eq:interpolant12}
\end{eqnarray}
whose leading terms at asymptotic $\Sigma$ are given in Eq.~(\ref{Eq:AsymPi}), whereas below the matching surface we set $\PiPS[\text{int 1,2}] = \bar{\Pi}_1^\text{PS-pole}$. In \ref{App:Convergence} we will show that these functions converge to the true $\PiPS$ in the limit $N\to\infty$ when matched to exact low-energy input using the convergence property of a Taylor series. The two different forms given in Eq.~(\ref{Eq:interpolant12}) will be used to estimate the sensitivity of our numerical results on the specific choice of interpolation between low and high energies.\footnote{We also considered multiplying the asymptotic expression in Eq.~(\ref{Eq:AsymPi}) by Pad\'e approximants in $\Sigma^{-1}$. Using up to 3 free parameters and fixing them in the way discussed below, however, leads to poles within the $a_\mu$ integration domain, where $\bar{\Pi}_1$ is known to be analytic.}

The coefficients $b_i(r,\phi)$ are fixed from the requirement that the $\PiPS[\text{int i}]$ have the same value and the same $N-1$ $\Sigma$-derivatives as the low-energy representation if evaluated at $\Sigma = \Sigma^\text{match}(r,\phi)$ for each $(r,\phi)$. No expansion is performed in $r$ and $\phi$, which is crucial to obtain a smooth transition to the low-energy regime.

Determining the optimal value of $N$ is a non-trivial issue. On the one hand, larger values of $N$ seem to be preferable since the true $\bar{\Pi}_1$ is analytic for space-like momenta and thus all derivatives are continuous. On the other hand, matching many derivatives leads to a function that is almost saturated by the low-energy input contribution up to considerably higher energies than $\Sigma^\text{match}(r,\phi)$. Since it is desirable to match at least one derivative in order to have $\bar{\Pi}_1$ differentiable at the matching point, we will use $N \in \{2,3\}$ in order to estimate the dependence on $N$.

Interpolation functions with a logarithmic dependence on $\Sigma$ are not forbidden.
This can stem, for example, from non-perturbative corrections leading to terms like $\ln{(Q_i^2/M^2)}$, where $M$ is some non-perturbative mass scale. In fact, the Regge model considered in Refs.~\cite{BernSDCShort,BernSDCLong} leads to interpolants containing terms like $Q^{-4}\ln{(Q^2/\sigma^2)}$ for $Q_i^2 = Q^2 \to \infty$, where $\sigma^2$ could {\it e.g.}\ be the Regge slope of the excited pseudoscalar masses. In order to allow for such a logarithmic approach of the asymptotic expression, we additionally consider the alternative interpolant
\begin{eqnarray}
\PiPS[\text{int 3}](\Sigma, r, \phi) & = & \PiPS[\text{asymp}](\Sigma, r, \phi) \nonumber\\
&&\times\left(1 + b_1(r,\phi) \Sigma^{-1} \ln\bigg(\frac{\Sigma}{\LQCD^2}\right) + \nonumber\\
&&\quad\sum_{i = 1}^{N-1}b_{i+1}(r, \phi) \Sigma^{-i}\bigg)
\label{Eq:interpolant3}
\end{eqnarray}
and use again $N \in \{2,3\}$.


\subsection{The matching surface \texorpdfstring{$\Sigma^\text{match}$}{Sigma match}}
\label{Sec:Sigmamatch}
The remaining crucial ingredient in our procedure is the function $\Sigma^\text{match}(r,\phi)$, which determines the value of $\Sigma$ at which the matching is performed for given $r$ and $\phi$. Choosing it too low leads to important modifications of $\PiPS$ at low energies with consequent overestimation of $\aPS$.\footnote{We denote by $\aPS$ the result of the integral in Eq.~(\ref{Eq:MasterFPi1}) for $\bar{\Pi}_1 \equiv \PiPS$.} Conversely, choosing $\Sigma^\text{match}$ too high assumes the low-energy input to dominate beyond what is expected according to mass and phase-space considerations and thus leads to underestimate $\aPS$.

For small values of $Q_3^2$, the $\pi^0$, $\eta$, $\eta'$ poles are assumed to dominate independently of $Q_{1,2}^2$, due to the pole at $Q_3^2 = -m_\text{PS}^2$ (see Secs.~\ref{Sec:SDCBTT} and~\ref{Sec:OPE}). This implies that no matching is needed in this regime, {\it i.e.}\ $\Sigma^\text{match}(1,\pi) = \infty$. The most general function that is analytic for all $(r,\phi)$ except for a (first-order) pole at $(r,\phi) = (1,\pi)$ can be written as
\begin{equation}
\Sigma^\text{match}(r,\phi) = \frac{3m^2}{1+r \cos \phi}\left(1 + P(r \cos \phi, r \sin \phi)\right)\, ,
\label{Eq:MatchSc}
\end{equation}
where $m^2$ determines the matching scale at $r = 0$ and $P$ is a polynomial with two arguments and no constant term. The transformation property of $\bar{\Pi}_1$ under crossing specified in Eq.~(\ref{Eq:crossing}) restricts $P$ to contain only even powers of $r \sin \phi$.

The parameter $m^2$ sets the absolute mass scale of $\Sigma^\text{match}$ and should thus be related to the masses of the states affecting $\bar{\Pi}_1$ beyond the ones explicitly included, namely $\pi^0$, $\eta$, $\eta'$ here.
In the following, we will assume that contributions to $\bar{\Pi}_1$ in the $g-2$ kinematics stemming from multi-particle intermediate states are dominated by narrow resonances while non-resonant effects lead to negligible corrections to the matching procedure and can be simply added to our final results.\footnote{We have checked the effects of the inclusion of the pion-loop contribution to $\bar{\Pi}_1$~\cite{BTT3} in the low-energy representation. Since the two-pion state contains a five-dimensional representation of the isospin group, a full decomposition into $\PiPS$ is not possible. However, even if its complete contribution is added to the isovector channel, we find that at the current level of accuracy it is irrelevant whether the pion loop is included in the matching procedure or not.} This is realized for example in the large-$N_c$ limit of pure QCD: since the short-distance expressions for $\bar{\Pi}_1$ in both symmetric and asymmetric limits scale like $N_c$, these can indeed be saturated by single-meson exchanges (see Ref.~\cite{deRafael}). Non-resonant contributions from multi-hadron intermediate states (like $2\pi$, $2 K$, $\pi\eta$, $3\pi$, \dots) are sub-leading for large $N_c$ and thus cannot contribute to the SDCs. Since scalar mesons have no impact on $\bar{\Pi}_1$, the lightest states beyond the ground-state pseudoscalars that are the most relevant at small $Q_3^2$ (see Sec.~\ref{Sec:SDCBTT}) are the axial mesons like $a_1(1260)$ and the tensor mesons like $f_2(1270)$, with masses in the \SIrange[range-units=single, range-phrase=--]{1}{2}{\GeV} region, whose effects on $a_\mu^\text{HLbL}$ can presently be estimated only using hadronic models.

For $P(x,y) = 0$, $\Sigma=\Sigma^\text{match}$ corresponds to $Q_3^2 = m^2$. Since a state of mass $M$ ceases to be suppressed by the denominator $(Q_3^2 + M^2)$ compared to lighter states when $Q_3^2$ approaches $M^2$, $m^2$ should be chosen well below $M^2$. At the same time, it should not be taken too small, because we do not expect any large contribution to $\bar{\Pi}_1$ at $Q_3^2 \ll M^2$. We thus regard $m^2 = \SI{0.5}{\,\GeV^2}$ as a good starting point for our analysis. In Sec.~\ref{Sec:NumMatch} we will discuss a range of choices for this parameter as well as the effects of the polynomial
\begin{equation}
P(x,y) = \sum_{i=0}^{M} \sum_{j=0}^{M} \frac{1}{i!\ j!} p_{i,j}\ x^i y^j\,,
\label{Eq:PolynomialSigmaMatch}
\end{equation}
which we have estimated by means of a Monte Carlo sampling over the coefficients $p_{i,j}$.
\section{Numerical results and error analysis}
\label{Sec:Num}
\subsection{The isovector channel}
\label{Sec:NumIV}
The isovector channel is best suited to our method since it is characterized by a large contribution from the low-energy region dominated by the well-known pion pole, which does not mix (strongly) into the other flavor channels. The lightest one-particle intermediate state beyond the $\pi^0$ in this channel is the $a_1(1260)$, whose effect at low energies is suppressed by the large mass gap. The numerical dominance of this channel at low energies, however, does not imply that the same holds true at intermediate and high energies. In fact, the values of $C_a$ in Eq.~(\ref{Eq:Ca}) make the flavor singlet channel the numerically most important one in the asymptotic region where meson masses can be neglected, {\it i.e.}\ for $Q_i^2 \gg \LQCD^2$. In Sec.~\ref{Sec:NumIS} we will discuss the inclusion of $\eta/\eta'$ and in Sec.~\ref{Sec:NumAxials} also the case of the isovector ground-state axial, which is however affected by a larger degree of model dependence.

We start by selecting a \enquote{reference} set of assumptions and input parameters. The impact of their modifications will be assessed in the next sections and will define the range of our predictions in the form of an uncertainty band. This procedure allows us also to examine how the estimate of the effects of SDCs would be improved by more precise information on the pion pole, the contributions from states with masses around \SI{1}{\GeV} and the asymptotic regime.

\begin{figure*}
	\centering
	\includegraphics{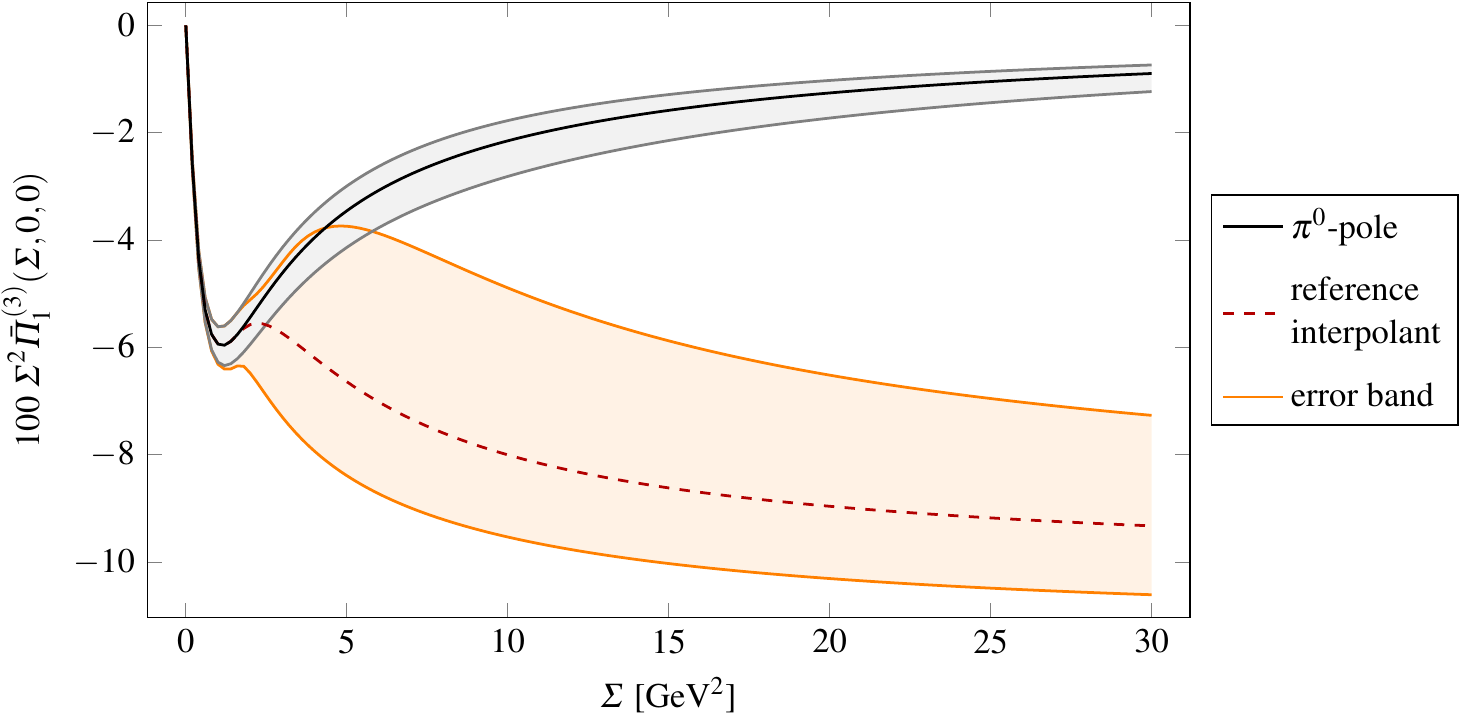}

	\caption{The pion pole contribution and associated uncertainty from Refs.~\cite{PionTFFshort,PionTFF} vs. the reference interpolant and its error band which includes all sources of uncertainty considered in the present analysis (see discussion in Secs.~\ref{Sec:NumTFF} -- \ref{Sec:NumMatch} below).}
	\label{Fig:matchPionCenter}
\end{figure*}

As low-energy reference input, we took the leading-order dispersive $\pi^0$ singly- and doubly-virtual TFFs~\cite{PionTFFshort,PionTFF}, while the corresponding $\Order{(\alpha_s)}$ correction is included in the uncertainty. As reference interpolating function, we used $\Pipion[\text{int 1}]$ with $N=3$ (see Eq.~(\ref{Eq:interpolant12})), which turned out to yield results that are central in the range spanned by the interpolants 1, 2 and 3 and $N\in \{2,3\}$ (cf.\ Eq.~(\ref{Eq:NumInterpolant}) below). For the asymptotic function, we included information from pQCD away from $r=0$ in the way explained in Sec.~\ref{Sec:HEInt}, while $\alpha_s$ corrections contribute to the uncertainty. For the matching surface we used Eq.~(\ref{Eq:MatchSc}) with $P(x,y) = 0$ and $m_\text{ref}^2 = \SI{0.5}{\GeV^2}$. The resulting function, which we call $\Pipion[\text{ref}]$, is shown in Fig.~\ref{Fig:matchPionCenter} for $r=0$ together with the uncertainty band for the interpolants that we are going to discuss in the next sections. Our reference outcome for the contribution to $\aLon$ due to the longitudinal SDCs in the isovector channel is
\begin{equation}
\Delta \aPion[\text{ref}] = \aPion[\text{ref}] - a_{\mu,\text{disp}}^{\pi^0\text{-pole}} = \num{2.56e-11}\,,
\label{Eq:refRes}
\end{equation}
where $\aPion[\text{ref}]$ comes from using $\Pipion[\text{ref}]$ in the master formula Eq.~(\ref{Eq:MasterFPi1}), and 
$a_{\mu,\text{disp}}^{\pi^0\text{-pole}}$ is given in Eq.~(\ref{Eq:aPionPole}) according to Refs.~\cite{PionTFFshort,PionTFF}.
In Sec.~\ref{Sec:ResultsIV} we will argue that our final result does not strongly depend on the choice of the reference set of parameters.

\subsubsection{Pion TFF uncertainties}
\label{Sec:NumTFF}

We shall now describe the effects of modifying the different ingredients of the reference configuration, one by one, starting from the pion TFF.
By propagating the errors quoted in Refs.~\cite{PionTFFshort,PionTFF} for the dispersive determination of the pion TFF and by summing the different contributions in quadrature, taking as well into account that a modification of the TFF affects both terms in Eq.~(\ref{Eq:refRes}), we obtained an asymmetric error band around the reference result with boundary values
\begin{equation}
\delta^+_\text{TFF} \Delta \aPion = \num{0.06e-11}\,, \quad \delta^-_\text{TFF} \Delta \aPion = \num{0.13e-11}\,,
\end{equation}
which correspond to the asymmetric error for the dispersive $\pi^0$ TFF. Given the smallness of these uncertainties, the (negative) correlation between them and the uncertainties of $a_{\mu,\text{disp}}^{\pi^0\text{-pole}}$ can be safely neglected.

In order to study the impact of different pion TFF parameterizations, we compared the previous results against the ones obtained using, both for the construction of the interpolant and the evaluation of $a_{\mu}^{\pi^0\text{-pole}}$, the $C^1_2$ Canterbury approximant with $a_{\pi; 1,1} = 2 b_\pi^2$ of Ref.~\cite{Canterbury}  and the Dyson-Schwinger TFF from Ref.~\cite{DSE}. We obtained
\begin{eqnarray}
\Delta \aPion[\text{Can}] &=& \num{2.60e-11}\,, \nonumber\\ \Delta \aPion[\text{DSE}] &=& \num{2.52e-11}\,, 
\end{eqnarray}
which are both compatible with the reference result within the range given above. We conclude that the outcome of our analysis is very robust against changes in the TFF input and that the present knowledge of the pion TFF is sufficient for our purposes.

\subsubsection{Asymptotic uncertainties}
\label{Sec:AsymUnc}
Here we focus on the uncertainties in $\Pipion[\text{asymp}]$ (see Eq.~(\ref{Eq:AsymPi})), which are related to
\begin{itemize}
	\item the choices made in the fit to the quark-loop result that lead to Eq.~(\ref{Eq:AsymPi_a_pion}), namely the degree $M$ of the polynomial and the radius $r_\text{max}$ of the fitting domain;
	\item $\alpha_s$ corrections to the OPE constraint as given by Eq.~(\ref{Eq:OPEConstraintNLO});
	\item $\alpha_s$ corrections to the quark loop. 
\end{itemize} 
We start by discussing the fit to the quark-loop result. In Sec.~\ref{Sec:HEInt}, we chose $M=2$, which leads to a strongly improved fit quality compared to $M=1$. Considering a larger value of $M$ gives an estimate of the errors made by approximating the pQCD result by a polynomial at $r < r_\text{max}$ at fixed asymptotic $\Sigma$ and by extrapolating to the regime $r>r_\text{max}$, which is unknown except for the OPE constraint. Choosing $M=3$ shifts the result for $\Delta \aPion$ by only \num{0.02e-11} indicating that the truncation at $M=2$ is sufficient. We also studied the effects of a substantial reduction of the radius, namely from $r_\text{max} = 0.9$ down to $r_\text{max} = 0.5$, where no logarithm of ratios of squared momenta is larger than 1. We found that this leads to a small shift (\num{0.07e-11}). We did not consider $r_\text{max}>0.9$ since fixed-order pQCD is not expected to converge for $r$ close to 1 due to large logarithms. Combining linearly the uncertainties from the choice of $M$ and the fitting radius gives 
\begin{equation}
\delta_\text{pQCD fit} \Delta \aPion = \num{0.09e-11}\,,
\end{equation}
with respect to the reference contribution of longitudinal SDCs to the pion pole input in Eq.~(\ref{Eq:refRes}).

Let us now focus on the estimate of the separate perturbative corrections to either the OPE or the pQCD result. Since those concerning the OPE should not be extrapolated into the domain of validity of pQCD, for asymptotic $\Sigma$ we write (cf.\ Eq.~(\ref{Eq:OPEConstraintNLO}))
\begin{eqnarray}
\Pipion[\text{asymp, $\delta$OPE}] & = & \Pipion[\text{asymp}]\bigg[1-\frac{\alpha_s(\mu^2=Q_1^2 + Q_2^2)}{\pi} \nonumber \\
&&\qquad\times \operatorname{\theta}\!\left(A-\frac{Q_3^2}{Q_1^2 + Q_2^2}\right)\bigg]\,.
\label{Eq:Pi1NLOMVOPE}
\end{eqnarray}
Here the Heaviside step function $\theta$ ensures that the perturbative correction only affects a region around $Q_3^2 = 0$, whose size can be varied via the free parameter $A$. By setting $A=1/29$, this region does not extend into the $r<0.9$ domain. The choice of the renormalization scale $\mu^2$ is the same as in Ref.~\cite{PionTFF} and agrees with our discussion in Sec.~\ref{Sec:MVpert} up to a factor of 2 since $-\hat{q}^2 = (Q_1^2 + Q_2^2)/2 - Q_3^2/4 \approx (Q_1^2 + Q_2^2)/2$ in the relevant regime. In our numerical analysis, for the running of $\alpha_s$ we used the three-flavor one-loop beta function and matched to $\alpha_s(\mu^2 = M_\tau^2) = 0.35$.

According to the discussion in Sec.~\ref{Sec:SDC}, the OPE constraint in the chiral limit is saturated by the pion pole at $Q_3^2 = 0$ to all orders in perturbation theory. For this reason in a consistent analysis the OPE coefficient and the pion TFF in the symmetric limit should be taken at the same perturbative accuracy. Hence we replaced in Eq.~(\ref{Eq:interpolant12}) $\Pipion[\text{asymp}]$ by $\Pipion[\text{asymp, $\delta$OPE}]$ and matched the correspondingly modified $\Pipion[\text{int 1}]$ to the pion-pole contribution with TFFs including $\Order(\alpha_s)$ effects~\cite{Braaten, PionTFF}.\footnote{We thank Bai-Long Hoid for kindly providing us with a numerical representation of the dispersive pion TFF with $\Order(\alpha_s)$ corrections.} Using this interpolant and this pion-pole result, our outcome for $\Delta \aPion$ is larger than the reference result Eq.~(\ref{Eq:refRes}) by
\begin{equation}
\delta_\text{NLO OPE}^+ \Delta \aPion = \num{0.01e-11}\,.
\label{Eq:uncNLOMV}
\end{equation}
For $A=1/3$, the domain where the correction applies extends down to $r=0.25$, but nevertheless the shift of $\Delta \aPion$ turns out to be \num{-0.05e-11} and thus still negligible. The smallness of these shifts can be understood from the large values of $\Sigma^\text{match}$ in the region where these perturbative corrections apply. For this reason, the effect is almost completely included in the pion pole contribution, where it also has a small impact~\cite{PionTFF}.

Since the NLO calculation of the quark loop has not been performed yet, we can only provide a rough estimate of the uncertainty related to unknown ${\cal O}(\alpha_s)$ corrections. We assumed in analogy with Eq.~(\ref{Eq:Pi1NLOMVOPE}),
\begin{eqnarray}
\Pipion[\text{asymp, $\delta$pQCD}] & = & \Pipion[\text{asymp}] \bigg[1-\frac{\alpha_s(\mu^2=\Sigma)}{\pi} \nonumber\\
&&\qquad\times\theta\left(r_\text{max} - r\right)\bigg]\,,
\label{Eq:Pi1NLOpQCD}
\end{eqnarray}
and as in the leading-order quark loop fit, we set $r_\text{max}=0.9$. Using this expression in Eq.~(\ref{Eq:interpolant12}) for the matching to the pion-pole with leading-order dispersive TFF, we obtained a shift of \num{-0.18e-11} compared to the reference result. Even when inflating this uncertainty by a factor of 2,
\begin{equation}
\delta_\text{NLO pQCD} \Delta \aPion = \num{0.36e-11}
\end{equation}
this effect is still sufficiently small compared to the current precision goal. We stress that once NLO calculations become available, $\Pipion[\text{asymp}]$ should be constructed to \emph{analytically} interpolate between the NLO expressions for the OPE and the quark loop. The discontinuous functions employed here only serve to provide a ballpark estimate of NLO effects.


\subsubsection{Choice of interpolation functions}
\label{Sec:NumInt}
In Eqs.~(\ref{Eq:interpolant12}) and (\ref{Eq:interpolant3}) we have introduced three different interpolation functions, characterized by two or three free parameters to be matched to the low-energy representation. The corresponding results for the contribution from longitudinal SCDs are
\begin{eqnarray}
\Delta \aPion[\text{int 1},\ N=2] & = & \num{3.18e-11}\,, \nonumber \\
\Delta \aPion[\text{int 1},\ N=3] & = & \num{2.56e-11}\,, \nonumber \\
\Delta \aPion[\text{int 2},\ N=2] & = & \num{2.75e-11}\,, \nonumber \\
\Delta \aPion[\text{int 2},\ N=3] & = & \num{2.16e-11}\,, \nonumber \\
\Delta \aPion[\text{int 3},\ N=2] & = & \num{2.69e-11}\,, \nonumber \\
\Delta \aPion[\text{int 3},\ N=3] & = & \num{1.94e-11}\,,
\label{Eq:NumInterpolant}
\end{eqnarray}
where $\Delta \aPion[\text{int 1},\ N=3] = \Delta \aPion[\text{ref}]$ given by Eq.~(\ref{Eq:refRes}) has been included for completeness.

We observe that the slower logarithmic approach to the asymptotic limits in the interpolant 3 leads to smaller results, especially when compared to the similar interpolant 1. Setting
\begin{equation}
\delta_\text{int} \Delta \aPion = \num{0.62e-11}\,,
\end{equation}
all values listed above are within the range $\Delta \aPion[\text{ref}] \pm \delta_\text{int} \Delta \aPion$.


\subsubsection{Choice of \texorpdfstring{$\Sigma^\text{match}(r,\phi)$}{sigma match}}
\label{Sec:NumMatch}
The function $\Sigma^\text{match}(r,\phi)$ in Eq.~(\ref{Eq:MatchSc}) contains the mass parameter $m$ 
and the polynomial $P(x,y)$, which has been set equal to zero so far. 
We have argued in Sec.~\ref{Sec:Sigmamatch} that $m$ should be chosen considerably smaller than the mass $M$ of the lightest resonances contributing to $\bar{\Pi}_1$ in addition to the ground-state pseudoscalar mesons. For this reason, for the reference interpolant we set $m^2=\SI{0.5}{\GeV^2}$. Here we discuss the effects of alternative choices for this parameter within a range between $m_\text{min}$ and $m_\text{max}$. 

Since according to Sec.~\ref{Sec:Sigmamatch} a conservative choice for the upper end of the range is $m_\text{max} \simeq M$, we set $m_\text{max}^2 = \SI{1}{\GeV^2}$. In order to determine an appropriate value for $m_\text{min}$, one has to estimate isovector contributions beyond the $\pi^0$-pole. Following our argument in Sec.~\ref{Sec:Sigmamatch}, it is sufficient to restrict ourselves to single-particle intermediate states and focus on the one giving the largest effect at energies around $m$. We assumed this to be given by the pseudoscalar $\pi(1300)$ for the following reasons. Models for tensor mesons around \SI{1}{\GeV} give similar or smaller contributions to $a_\mu^\text{HLbL}$ \cite{Pauk:2014rta, Danilkin:2016hnh, Danilkin:2019mhd}.
For ground-state axials, recent studies based on different approximations and hadronic models yield quite different numerical results, see {\it e.g.}\ Refs.~\cite{Pauk:2014rta,Roig:2019reh,HolographyVienna,HolographyItaly}, leading to large uncertainties. If future model-independent analyses show that axial-meson exchanges are responsible for significant effects in $\Pipion$ also at relatively small momenta, then these contributions should be added to the pion pole before the matching is performed since our procedure relies on a sufficiently precise knowledge of $\Pipion$ below $\Sigma^\text{match}(r,\phi)$. Neglecting issues related to model dependence, in Sec.~\ref{Sec:NumAxials} we will discuss the inclusion in our procedure of information from holographic QCD on the lightest axial meson.

\begin{figure*}
	\centering
	\includegraphics{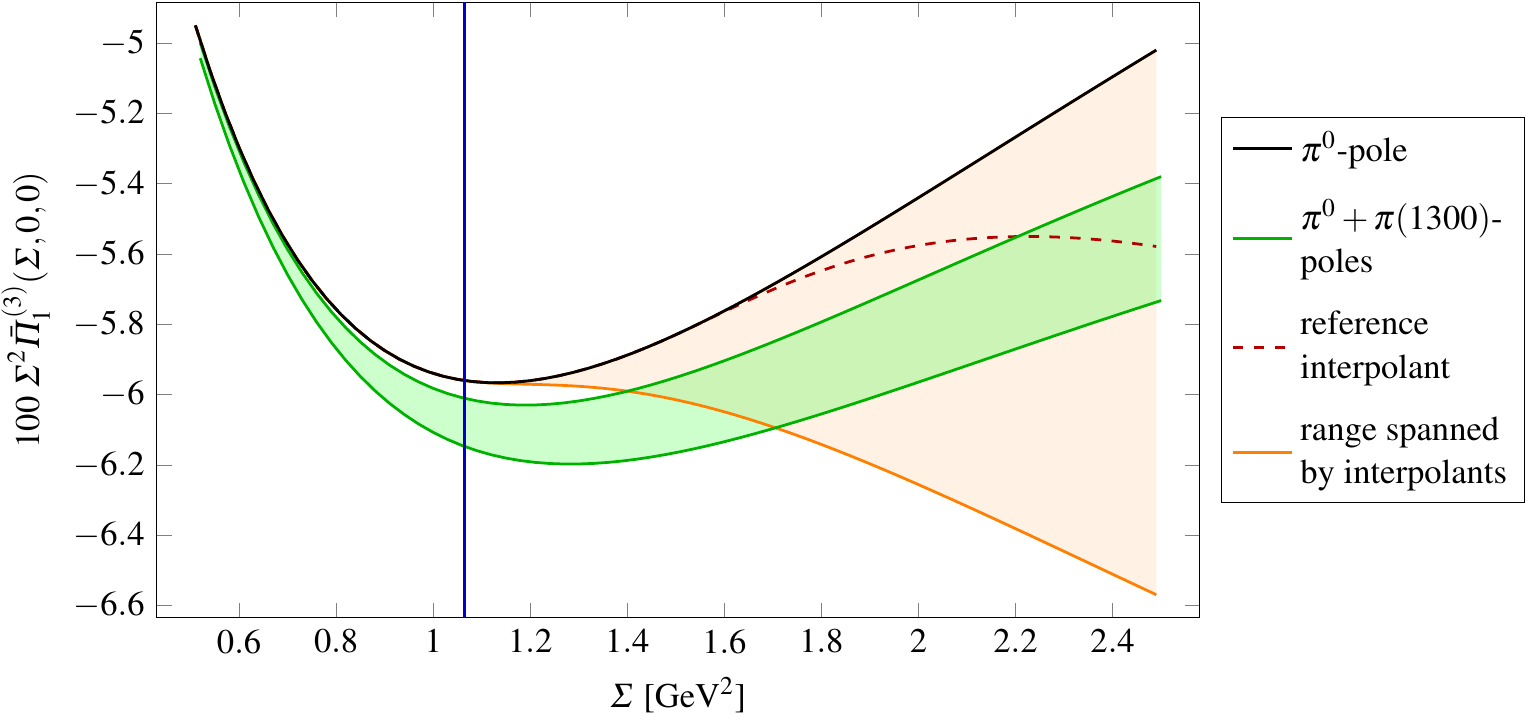}

	\caption{The figure displays the dispersive pion pole contribution, the reference interpolant and the (orange) band corresponding to the various choices of the parameter $m$. The blue line indicates the value of the matching surface for $m^2 = m_\text{min}^2$. The green band shows the sum of the $\pi^0$- and $\pi(1300)$-pole contributions, where the latter has been calculated using input from R$\chi$T and phenomenology, including errors.}
	\label{Fig:mFromRChiT}
\end{figure*}

As for the light pseudoscalars, the interaction of $\pi(1300)$ with two photons can be described by a TFF, which determines the contribution to $\bar{\Pi}_1$ as in Eq.~(\ref{Eq:Pi1PSpole}).
In our analysis we used as input the $\pi(1300)$ TFF derived in Ref.~\cite{KampfRChiT} in the framework of Resonance Chiral Theory (R$\chi$T)~\cite{RChiT}. We fixed the free parameters in Eq.~(69) of Ref.~\cite{KampfRChiT} by requiring that (i) the $\pi^0$-TFF satisfies the Brodsky-Lepage condition, {\it i.e.}\ $F_{\pi^0\gamma^*\gamma^*}(-Q^2,0) = 2 F_\pi/Q^2 +\Order(Q^{-4})$ and (ii) the two-real-photon limit of the excited pion TFF is in the range $F_{\pi(1300)\gamma^*\gamma^*}(0,0)\in[0,\num{0.0544}]\si{\GeV^{-1}}$, argued for in Ref.~\cite{BernSDCLong} based on experimental results~\cite{Acciarri:1997rb,Salvini:2004gz}. The upper boundary of this interval leads to the most conservative error estimate in our analysis, and is used in the following. 

Our procedure to determine $m_\text{min}$ can be illustrated by means of Fig.~\ref{Fig:mFromRChiT}. Given a value of $m_\text{min}$, at large enough $\Sigma$, the range of interpolants (orange band) spanned by $m\in [m_\text{min},m_\text{max}]$ safely includes the green band representing the sum of the contributions from $\pi^0$ and $\pi(1300)$, including errors on the latter due to the range for $F_{\pi(1300)\gamma^*\gamma^*}(0,0)$. Since this is not the case at small $\Sigma$, the contribution to $a_\mu^\text{long}$ from this region is underestimated in our approach. To gauge this effect, we calculated the integral in Eq.~(\ref{Eq:MasterFPi1}) with
\begin{equation}
\bar{\Pi}_1 = \bar{\Pi}_1^\text{$\pi^0$-pole} + \bar{\Pi}_1^\text{$\pi(1300)$-pole} - \Pipion[\text{int}](m=m_\text{min})
\label{Eq:Pi_miss}
\end{equation}
using the maximal $\pi(1300)$ contribution and restricting the $\Sigma$-domain to the region below the point where the bands start to fully overlap (as a function of $r$ and $\phi$). This integral gives the missed contribution $\aPion[\text{missed}]$ at a fixed $m_\text{min}$. Repeating this calculation for different values of $m_\text{min}$ yields the function $\aPion[\text{missed}](m_\text{min})$ and by inverting this, we determined $m_\text{min}$ by fixing $\aPion[\text{missed}]$ to values well below the accuracy goal set by forthcoming experimental results.
For $\aPion[\text{missed}] = \num{0.5e-11}$ we obtained $m_\text{min, 1}^2 = \SI{0.35}{\GeV^2}$ and for $\aPion[\text{missed}] = \num{0.2e-11}$, $m_\text{min, 2}^2 = \SI{0.13}{\GeV^2}$.

Numerically, $m_\text{max} = \SI{1}{\GeV}$ leads to the shift $\delta_{m}^- \Delta \aPion = \num{1.20e-11}$ and the two values $m_\text{min, 1}$ and $m_\text{min, 2}$ yield $\delta_{m, 1}^{+\prime} \Delta \aPion = \num{0.81e-11}$ and $\delta_{m, 2}^{+\prime} \Delta \aPion = \num{3.66e-11}$, respectively. If we add $\aPion[\text{missed}](m_\text{min, \{1, 2\}})$ to the latter numbers, we obtain the conservative estimates
\begin{eqnarray}
\delta_{m}^- \Delta \aPion & = & \num{1.20e-11}\,, \nonumber \\
\delta_{m, 1}^+ \Delta \aPion & = & \num{1.31e-11}\,, \  \delta_{m, 2}^+ \Delta \aPion = \num{3.86e-11}\,.
\label{Eq:NumMatching}
\end{eqnarray}
In the following we will use  $\delta_{m, 1}^+ \Delta \aPion$ for the main results and keep $\delta_{m, 2}^+ \Delta \aPion$ as an alternative, even more conservative uncertainty. 

We also considered a different parameterization for the $\pi(1300)$ TFF, namely the one given by the Regge model in Refs.~\cite{BernSDCShort,BernSDCLong}. Using the empirical $m_{\pi(1300)} = \SI{1.30}{\GeV}$ instead of the Regge-model value of \SI{1.36}{\GeV} used in these references and following the same procedure discussed above, we obtained $m_\text{min, 1}^2 = \SI{0.53}{\GeV^2}$ and $m_\text{min, 2}^2 = \SI{0.20}{\GeV^2}$. This leads to
\begin{equation}
\delta_{m, 1}^+ \Delta \aPion = \num{0.36e-11}\, ,\quad \delta_{m, 2}^+ \Delta \aPion = \num{2.63e-11}\,,
\end{equation}
where we again added $a_\mu^\text{missed}$ to the uncertainties in the upward direction. For our final result we use the more conservative uncertainty estimates given in Eq.~(\ref{Eq:NumMatching}).

In order to study the effects of the polynomial in $\Sigma^\text{match}(r,\phi)$, Eq.~(\ref{Eq:PolynomialSigmaMatch}), we set $M=2$ and sampled the free parameters according to a standard normal distribution. Since the pion pole gives an excellent approximation of $\bar{\Pi}_1$ for very small $\Sigma$ at any $(r,\phi)$, we only allowed for parameters giving $\Sigma^\text{match}(r,\phi)>\Sigma_t$ for all $(r,\phi)$, where $\Sigma_t$ is defined as the smallest value of $\Sigma$ such that
\begin{equation}
\frac{\bar{\Pi}_1^\text{$\pi(1300)$-pole}(\Sigma,r,\phi)}{\bar{\Pi}_1^\text{$\pi^0$-pole}(\Sigma,r,\phi)} = 0.02
\label{Eq:MCRatio}
\end{equation}
holds for some $(r,\phi)$. With $\bar{\Pi}_1^\text{$\pi(1300)$-pole}$ calculated using R$\chi$T, we obtained $\Sigma_t = \SI{0.57}{\GeV^2}$. This condition ensures that there are no large contributions from our interpolation at points where R$\chi$T predicts a very small excited pion contribution. From this we calculated a distribution of results for $\Delta \aPion$, which features a Gaussian-like peak close to the reference result and asymmetric tails, and read off the \SI{16}{\percent} quantiles from both sides corresponding to the $1\sigma$ errors for a Gaussian. This gives
\begin{minipage}{\columnwidth}
\begin{equation}
\delta_{P(x,y)}^+ \Delta \aPion = \num{0.39 e-11}\ ,\quad \delta_{P(x,y)}^- \Delta \aPion = \num{0.32 e-11}\,.
\end{equation}
\end{minipage}
We have checked that this result is stable against the inclusion of terms of order 3 in the polynomial and moderate changes in the value of the ratio in Eq.~(\ref{Eq:MCRatio}).


\subsubsection{Estimate of the effects of longitudinal SDCs in the isovector channel}
\label{Sec:ResultsIV}

\begin{table}
	\caption{The effects on $a_\mu^{\text{HLbL}}$ of longitudinal SDCs assuming that the low-energy region is dominated by ground-state pseudoscalar poles, whose contributions are taken as input. In each flavor channel the results are presented as the shifts $\Delta a_{\mu,\text{ref}}$ with respect to the pole contributions for a specific reference set of parameters and a list of uncertainties corresponding to different choices for each of these parameters. In the last two rows, these uncertainties are added in quadrature and the final range is symmetrized. See main text for details.}
	\label{Tab:results}
	{\def\arraystretch{1.5}\tabcolsep=10pt
		\begin{tabular*}{\columnwidth}{@{\extracolsep{\fill}}lrrr@{}}
			\hline
			& $\pi^0$ & $\eta$ & $\eta'$ \\ \hline
			$\Delta a_{\mu,\text{ref}} \times 10^{11}$ & $2.56$ & $2.58$ & $3.91$ \\ \hline
			$\delta_\text{TFF} \Delta a_{\mu} \times 10^{11}$ & $~^{+0.06}_{-0.13}$ & $0.47$ & $0.30$ \\
			$\delta_\text{pQCD fit} \Delta a_\mu \times 10^{11}$ & $0.09$ & $0.08$ & $0.14$ \\
			$\delta_\text{NLO OPE} \Delta a_{\mu} \times 10^{11}$ & $~^{+0.01}_{-0.00}$ & $~^{+0.01}_{-0.00}$ & $~^{+0.02}_{-0.00}$ \\
			$\delta_\text{NLO pQCD} \Delta a_{\mu} \times 10^{11}$ & $0.36$ & $0.36$ & $0.55$ \\
			$\delta_\text{int} \Delta a_{\mu} \times 10^{11}$ & $0.62$ & $~^{+0.61}_{-0.65}$ & $~^{+0.74}_{-0.84}$ \\
			$\delta_{m, 1} \Delta a_{\mu} \times 10^{11}$ & $~^{+1.31}_{-1.20}$ & $~^{+1.27}_{-1.17}$ & $~^{+1.68}_{-1.60}$ \\
			$\delta_{P(x,y)} \Delta a_{\mu} \times 10^{11}$ & $~^{+0.39}_{-0.32}$ & $~^{+0.31}_{-0.33}$ & $~^{+0.32}_{-0.43}$ \\ \hline
			$\delta_\text{tot} \Delta a_{\mu} \times 10^{11}$ & $~^{+1.55}_{-1.44}$ & $~^{+1.56}_{-1.50}$ & $1.97$ \\
			$\Delta a_{\mu} \times 10^{11}$ & $2.6\pm 1.5$ & $2.6\pm 1.5$ & $3.9\pm 2.0$ \\ \hline
		\end{tabular*}
	}
\end{table}


\begin{figure}
	\centering
	\includegraphics{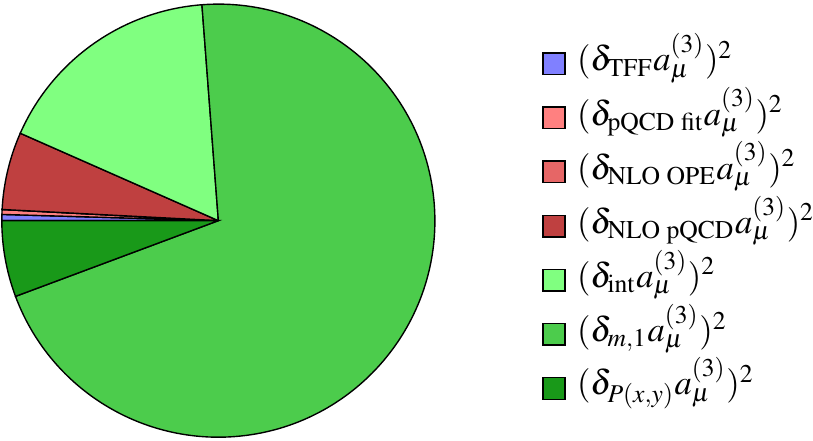}
	
	\caption{Relative contributions to the total uncertainty in the isovector channel. For asymmetric errors the mean of the squared errors is used.}
	\label{Fig:PieChartErrors}
\end{figure}


The $\pi^0$-column of Tab.~\ref{Tab:results} collects all uncertainties in our estimate of the effects of longitudinal SDCs in the isovector channel, as described in the previous subsections. By combining them in quadrature we get
\begin{equation}
\delta_\text{tot}^+ \Delta \aPion = \num{1.55e-11}\,, \quad \delta_\text{tot}^- \Delta \aPion = \num{1.44e-11}\,.
\end{equation}
Since we do not regard the reference parameterization as the central value, we symmetrized the uncertainty to finally obtain the range
\begin{equation}
\Delta \aPion = \num[separate-uncertainty]{2.6\pm 1.5 e-11}\, .
\end{equation}
Using instead $\delta_{m, 2}^+ \Delta \aPion$ in Eq.~(\ref{Eq:NumMatching}), the final result would be \num[separate-uncertainty]{3.8 \pm 2.7e-11}. Notice that, despite the fact that it likely overestimates the range of longitudinal short-distance effects, this interval is still definitely compatible with the current precision goal.

Fig.~\ref{Fig:PieChartErrors} shows the contributions to the quadratic error from the different sources discussed above. The vastly dominant effect stems from the interpolation between low and high energies, with an especially crucial role played by the choice of $m$, the scale at which the matching between the low-energy representation of $\bar{\Pi}_1$ and the interpolant is performed. The uncertainties $\delta_\text{int}$, $\delta_m$ and $\delta_{P(x,y)}$ could be reduced by additional low-energy input concerning further intermediate states and higher-order terms in the symmetric and asymmetric OPEs, which would help constrain the coefficients $b_i(r,\phi)$ in the interpolants in Eqs.~(\ref{Eq:interpolant12}) and (\ref{Eq:interpolant3}). The uncertainties related to the perturbative corrections are considerably smaller. While we do not expect that calculations of $\alpha_s$ corrections will crucially improve the final estimate, these perturbative results will definitely be important to better assess the regime of validity of the asymptotic constraints and thereby verify and sharpen some of our assumptions. 

We have also checked that our results are robust against the choice of different reference sets of parameters. For example, if we set $m_\text{ref}$ equal to the previous boundary values for the uncertainty in the reference configuration, namely $m_\text{ref}^2 = \SI{0.35}{\GeV^2}$ and $m_\text{ref}^2 = \SI{1}{\GeV^2}$ and choose $m^2\in [0.35, 1]\si{GeV^2}$ as the range for the error estimation as before, we get
\begin{eqnarray}
\Delta \aPion[m_\text{ref}^2 = \SI{0.35}{\GeV^2}] & = & \num[separate-uncertainty]{2.8\pm 1.7e-11}\,, \nonumber \\
\Delta \aPion[m_\text{ref}^2 = \SI{1.00}{\GeV^2}] & = & \num[separate-uncertainty]{2.4\pm 1.5e-11}\,,
\end{eqnarray}
where all other sources of uncertainty are included. We obtained similar results by selecting as reference different interpolants or different values of the number of free parameters $N$ contained therein.


\subsection{The isoscalar contributions}
\label{Sec:NumIS}
In this section the procedure presented above for the isovector case is applied to the isoscalar channels with $\etaetap$-poles as low-energy input. In our analysis, we employed the Canterbury TFFs from Ref.~\cite{Canterbury} in the reference solution.\footnote{In the conventions of  Ref.~\cite{Canterbury}, we used the $C^1_2$ approximant with $a_{\etaetap;1,1} = 2 b_\etaetap^2$ as for the pion.} We determined the parameters encoding $\eta-\eta'$-mixing as explained in Secs.~\ref{Sec:OPE} and~\ref{Sec:pQCD} and obtained
\begin{equation}
C_\eta = \num{0.164}\,, \quad C_{\eta'} = \num{0.219}\,, \quad \delta_0 = \num{0.110}\,,
\label{Eq:CCan}
\end{equation}
which shows that $\delta_0$ is indeed sizable.

Following the same procedure for the construction of the reference interpolant as in Sec.~\ref{Sec:NumIV}, we found
\begin{equation}
\Delta a_{\mu, \text{ref}}^\eta = \num{2.58e-11}\,, \quad \Delta a_{\mu, \text{ref}}^{\eta'} = \num{3.91e-11}\,.
\end{equation}

The uncertainty estimation proceeds in the same way as in the isovector channel up to minor modifications. Since error bands for the doubly-virtual TFFs in all kinematic configurations are not available in the literature, we estimated uncertainties by considering another TFF representation, namely the one based on Dyson-Schwinger equations (DSE)~\cite{DSE}. This yields $\Delta a_{\mu, \text{ref}}^\eta = \num{2.11e-11}$ and $\Delta a_{\mu, \text{ref}}^{\eta'} = \num{4.20e-11}$. The fact that individual results for $\eta$ and $\eta'$ channels differ by \num{18} and \SI{8}{\percent}, but the sum only by \SI{3}{\percent} can be understood by comparing the mixing parameters
\begin{equation}
C_\eta^\text{DSE} = \num{0.148}\,, \quad C_{\eta'}^\text{DSE} = \num{0.228}\,, \quad \delta_0^\text{DSE} = \num{0.127}\,,
\label{Eq:CDSE}
\end{equation}
against those obtained from the Canterbury parameterization. These coefficients enter $\Pietaetap[\text{asymp}]$ quadratically, which leads to a reshuffling between $\aEta$ and $\aEtap$. Due to Eq.~(\ref{Eq:CPhysFlavor}) this effect drops out in the sum up to the anomaly correction affecting the OPE regime. As TFF contribution to the uncertainty on $\Delta \aEtaEtap$ we took the absolute value of the differences between the results from the Canterbury and DSE TFFs.

Since NLO results are not available for the $\etaetap$ TFFs, we estimated the NLO OPE uncertainty by simply rescaling the one in the pion channel by the ratio of the reference outcomes. Due to the smallness of this uncertainty, this is expected to be sufficiently accurate.

For the range $[m_\text{min}, m_\text{max}]$ and the minimal allowed value $\Sigma_t$ for $\Sigma^\text{match}$ in the Monte Carlo simulation for $P(x,y)$, we took the results from the isovector channel. We rescaled the $\pi(1300)$ term below the matching surface by the ratio of reference results when adding this contribution to $\delta_m \aEtaEtap$. This is justified by the fact that the first excited pseudoscalars in the three flavor channels have similar masses, despite the large mass difference of the pseudo-Goldstone bosons.

All results are collected in Tab.~\ref{Tab:results} and our final estimate for the longitudinal short-distance effects in $\aEtaEtap$ reads
\begin{eqnarray}
\Delta \aEta & = & \num[separate-uncertainty]{2.6 \pm 1.5 e-11}\,, \nonumber \\
\Delta \aEtap & = & \num[separate-uncertainty]{3.9 \pm 2.0 e-11}\, .
\end{eqnarray}
The relative contributions to the uncertainties are similar to the pion case illustrated in Fig.~\ref{Fig:PieChartErrors}. A more precise description of $\eta-\eta'$-mixing would of course help better separate the two isoscalar channels but would not play an important role in their sum leading to negligible shifts in the total contribution from longitudinal SDCs.


\subsection{Sum over the flavor channels and comparison with literature}
\label{Sec:NumResults}
Combining the results from Secs.~\ref{Sec:NumIV} and~\ref{Sec:NumIS}, obtained under the assumption that the ground-state pseudoscalar mesons dominate the low-energy region, our estimate for the total effect of the longitudinal SDCs on HLbL amounts to
\begin{eqnarray}
\Delta \aLon & = &
\Delta \aPion + \Delta \aEta + \Delta \aEtap \nonumber \\
& = & \num[separate-uncertainty]{9.1\pm 5.0 e-11}\, ,
\label{Eq:resultSum}
\end{eqnarray}
where we have combined the three uncertainties linearly since they originate from the same sources in all three channels. 

This result is remarkably close to what is expected based on flavor symmetry considerations. If the $U(3)$ symmetry emerging in the combined chiral and large-$N_c$ limit is assumed, then $\Delta \aEta + \Delta \aEtap = 3 \Delta \aPion$. Using our isovector uncertainty and adding linearly a standard \SI{30}{\percent} $U(3)$ breaking effect, we obtain
\begin{equation}
\Delta \aLon = \num[separate-uncertainty]{10.4 \pm 8.3e-11}\,.
\end{equation}
For this reason we do not expect that a more refined analysis of the subtler isosinglet contributions is going to change substantially our final results.

Refs.~\cite{BernSDCShort,BernSDCLong} have recently studied the possibility of saturating SDCs away from the chiral limit by including a tower of excited pseudoscalar states in the context of a Regge model matched to the pQCD quark loop. Their outcome is $\Delta \aLon = \num{13 \pm 6 e-11}$, which is well compatible with ours within errors. For the $\eta'$-channel, the Regge model yields $\Delta \aEtap = \num[separate-uncertainty]{6.5\pm 2.0 e-11}$, which is somewhat larger than our result but still compatible within errors.\footnote{The quoted result does not include the matching to the pQCD quark loop, which has only been performed for the sum of all channels in Refs.~\cite{BernSDCShort,BernSDCLong}.} This can partly be explained by the different value for $C_{\eta'}$ used in Refs.~\cite{BernSDCShort,BernSDCLong}, namely $C_{\eta'} = 0.239$, which results from imposing that Eq.~(\ref{Eq:CPhysFlavor}) holds exactly. 

\begin{figure*}
    \centering
    \includegraphics{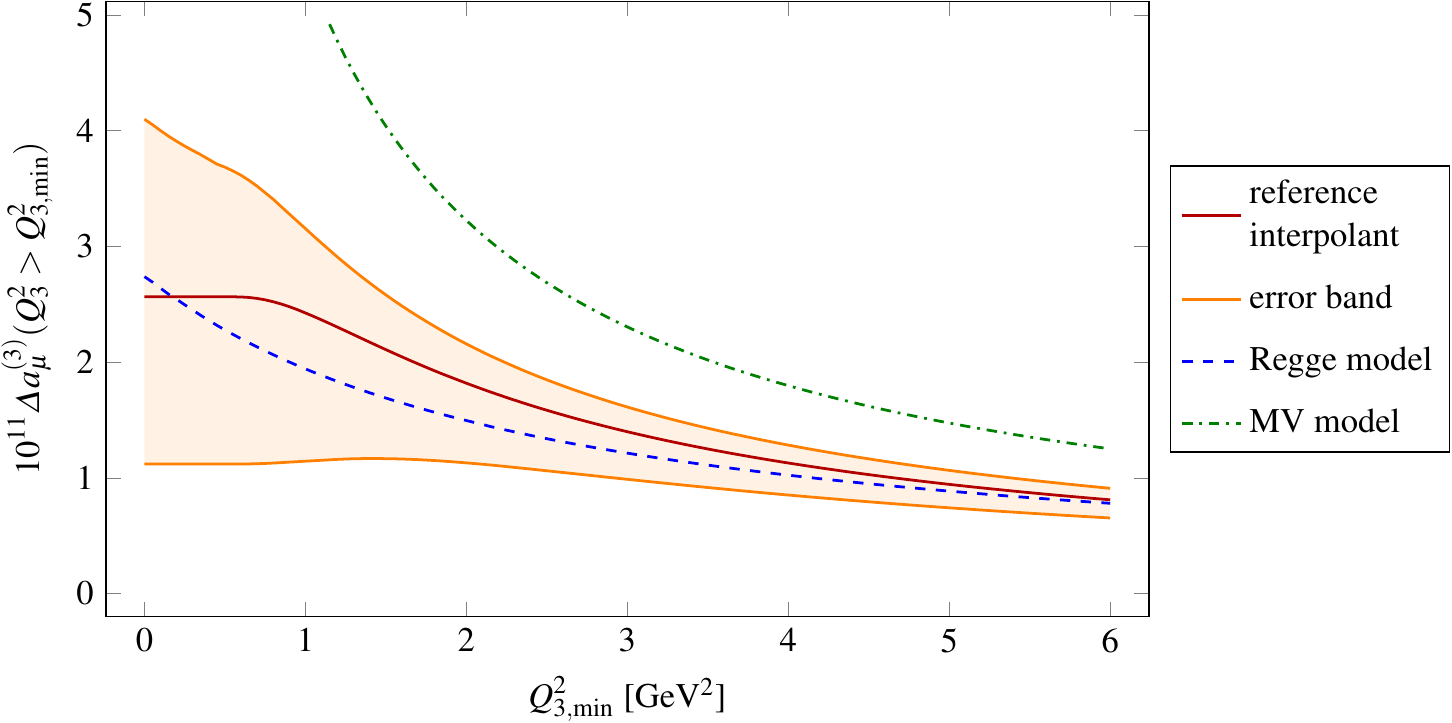}
    \caption{$\Delta \aPion$ as a function of a lower limit on $Q_3^2$ in Eq.~(\ref{Eq:MasterFPi1}): our reference result and corresponding error band against the tower of excited pseudoscalars in the large-$N_c$ Regge model 1 of Refs.~\cite{BernSDCShort,BernSDCLong} and the curve from the MV model~\cite{MV} evaluated using the up-to-date dispersive pion TFF. At small non-vanishing $Q_{3,\text{min}}^2$, our reference curve is constant due to the finite $\Sigma^\text{match}$, which for $P(x,y)=0$ corresponds to $m^2 = Q_3^2 = \text{const.}$, whereas the Regge model has a slope due to the absence of such a cutoff. The upper end of our error band shows a slope because of the inclusion of the $\pi(1300)$ contribution in that region.}
    \label{Fig:Q32MinPion}
\end{figure*}

Fig.~\ref{Fig:Q32MinPion} shows $\Delta \aPion$ as a function of a lower cutoff on $Q_3^2$ in our approach as well as the large-$N_c$ Regge model 1 of Refs.~\cite{BernSDCShort,BernSDCLong}. In order to obtain this plot, we calculated the integral in Eq.~(\ref{Eq:MasterFPi1}) as a function of a lower limit on $Q_3^2$ (which depends on $\Sigma$, $r$ and $\phi$) for both the full $\aPion$ as well as the pion pole contribution (cf.\ Eq.~(\ref{Eq:refRes})). The Regge model result lies within our error band for all $Q_{3,\text{min}}^2$.

Our estimate of longitudinal short-distance effects as well as the one in Refs.~\cite{BernSDCShort,BernSDCLong} are smaller than the shift obtained in Ref.~\cite{MV}, $\Delta \aLon = \num{23.5 e-11}$, which even increases to about \num{38e-11} if up-to-date TFF input is used~\cite{BernSDCLong}. These large values are due to two features of the MV model: the fact that the singly-virtual TFF is set to a constant over the whole integration region and not only in the OPE regime, and the fact that in the symmetric asymptotic limit the parametric momentum dependence is correct but its coefficient is too large. Both of these features can be clearly seen in Fig.~\ref{Fig:Q32MinPion} and are responsible for the discrepancies in the slope at small $Q_{3,\text{min}}^2$ and the values at large $Q_{3,\text{min}}^2$, respectively.

Refs.~\cite{HolographyVienna, HolographyItaly} have studied how the inclusion of an infinite tower of axial-vector mesons could help satisfy the OPE SDCs, focusing for this purpose on the relevant TFFs in the context of holographic QCD models. According to Ref.~\cite{HolographyVienna}, the tower of axial-vector mesons contributes \SIrange[range-phrase = --,scientific-notation = fixed, range-units = brackets, fixed-exponent = -11]{29e-11}{41e-11}{\noop} to $a_\mu^\text{HLbL}$ of which \SIrange[range-phrase = --, range-units = brackets]{57}{58}{\percent} are attributed to $\aLon$. Using instead holographic QCD input only for the momentum dependence of the TFF and fixing its normalization from experiment reduces the estimate of the contribution to $a_\mu^\text{HLbL}$ from the tower of axials to \num{22\pm5e-11}. Ref.~\cite{HolographyItaly} finds \num{14e-11} for the effect of axials on $\aLon$. Thus, the results of these studies appear to be at the high end of our range in Eq.~(\ref{Eq:resultSum}). However, we stress that comparing these numbers with our result is not properly justified. Indeed, while in these models the parametric $\Sigma$-dependences implied by pQCD and the OPE in the respective limits are correctly reproduced, the coefficients thereof are typically too small. In addition, the lightest multiplet of axials significantly alters $\bar{\Pi}_1$ at small photon virtualities, which implies that in our approach its contribution should be included in the low-energy representation. This aspect will be discussed in the next section, also to show how information on additional states in the \SI{1}{\GeV} region can be incorporated in our analysis.


\subsection{Including ground-state axial mesons at low energies}
\label{Sec:NumAxials}
Here we adopt a model-dependent approach to illustrate the application of our procedure to the case of the inclusion in the low-energy region of the lightest of the axial-vector mesons, for which no dispersive treatment in the BTT formalism is available yet. According to the holographic QCD models in Refs.~\cite{HolographyVienna, HolographyItaly} and using the notation of Ref.~\cite{HolographyVienna}, the contribution to $\bar{\Pi}_1$ of an axial meson of mass $M_A$ in the flavor channel $a$ can be written as
\begin{eqnarray}
\PiPS[\text{axial}] & = & -\frac{9 \,C_a^2}{16\pi^4 M_A^2}\left[Q_1^2 \,A(Q_1^2, Q_2^2) + Q_2^2 \,A(Q_2^2, Q_1^2)\right]\nonumber\\
&&\times A(Q_3^2,0)\,,
\label{Eq:Axial}
\end{eqnarray}
where $A(Q_1^2, Q_2^2)$ is the axial TFF.

Among the various scenarios analyzed in Ref.~\cite{HolographyVienna}, the hard-wall model by Hirn and Sanz (HW2)~\cite{Hirn:2005nr}, which was also studied in Ref.~\cite{HolographyItaly} with different parameters, reproduces best the measured mass, the measured equivalent two-photon decay width and the singly virtual momentum dependence measured by L3 for the lightest multiplet~\cite{Achard:2001uu, Achard:2007hm}. Furthermore, it yields asymptotic axial TFFs whose momentum dependence is consistent with the behavior derived in Ref.~\cite{Hoferichter:2020lap}. The infinite tower of axials has the correct momentum scaling in the asymmetric asymptotic regime dictated by the OPE constraint, but the coefficient is \SI{38}{\percent} too small~\cite{HolographyVienna}. 

We focused on the isovector channel, which is sufficient for our illustrative purposes, and thus on the inclusion of the $a_1$ meson. Based on the HW2 model, we obtained $a_{\mu,\text{HW2}}^{a_1} = \num{3.3e-11}$ for the $a_1$ contribution to $\aLon$. The rest of the tower of isovector axial mesons in this model yields $\Delta \aPion[A,\ \text{HW2}] = \num{0.8e-11}$, implying that in this framework about \SI{80}{\percent} of the total effect comes from the lightest state.\footnote{We thank Josef Leutgeb for checking these numbers and the ones for HW2(UV-fit) below.}

By matching the interpolant in Eq.~(\ref{Eq:interpolant12}) to the contributions from the pion and the holographic $a_1$ with the reference set of assumptions in Sec.~\ref{Sec:NumIV}, we obtained\footnote{We had to shift $C_\pi^2$ by \SI{1.7}{\percent} in order to account for the $a_1$ at small $Q_3^2$ and asymptotic $Q_1^2\sim Q_2^2$.}
\begin{equation}
\Delta \aPion[A] = \aPion[A] - a_{\mu,\text{disp}}^\text{$\pi^0$-pole} - a_{\mu,\text{HW2}}^{a_1} =\num{1.9e-11}\,.
\label{Eq:resAxialHW2}
\end{equation}
This result is more than twice as large as the resummed tower in HW2, $\Delta \aPion[A,\ \text{HW2}]$, but the significance of this discrepancy could only be assessed by a more sophisticated analysis including uncertainties, which is beyond the scope of this work. However, in the holographic model the infinite tower of axials does not fully saturate the pQCD nor the OPE constraints, which suggests that additional degrees of freedom besides axials should be included in a more realistic model.

We then considered the choice of parameters made in Ref.~\cite{HolographyItaly} and referred to as HW2(UV-fit) in Ref.~\cite{HolographyVienna}. This model is constructed to obey the OPE constraint exactly, but fails to describe low-energy physics like the $\rho$-meson mass, the pion TFF and the axial TFFs measured by L3. The longitudinal contribution from $a_1$ in this case amounts to $a_{\mu,\ \text{HW2(UV-fit)}}^{a_1} = \num{3.4e-11}$ and the tower of states increases the value by $\Delta \aPion[A, \text{HW2(UV-fit)}] = \num{0.8e-11}$. Our reference interpolant leads to $\Delta \aPion[A] = \num{1.4e-11}$, which is again larger than the model result. However, also in HW2(UV-fit) the pQCD constraint is not fully fulfilled by the tower of axials.

Neglecting issues related to intrinsic model dependence in the low-energy input, our method based on interpolants that by construction satisfy all constraints indicates that the effects of longitudinal SDCs are relatively small compared to the dominant low-energy contributions, and what is crucial in order to achieve higher precision is to gain control over the latter. We stress that a reliable prediction with a robust uncertainty estimate of the effects of axial meson exchanges would require model-independent input information.
\section{Conclusions}
\label{Sec:Con}
In this paper we have introduced a novel approach to incorporate longitudinal SDCs into the calculation of the HLbL contribution to the muon $g-2$. At variance with the previous estimates based on hadronic models, we have constructed general functions interpolating between low-, mixed- and high-energy regions, without resorting to specify which and how hadronic intermediate states are responsible for saturating the constraints. Furthermore, our method allows us also to study in detail the role played by parameters and assumptions in a transparent and numerically efficient way.

Our main premise is that an accurate low-energy representation of the longitudinal function $\bar{\Pi}_1(Q_1^2, Q_2^2, Q_3^2)$ entering the HLbL integral can be obtained by taking into account only intermediate states that are under good theoretical and numerical control. For the $\pi^0$, due to the location of its pole, the form of this low-energy representation can be straightforwardly extended even to large $Q_1^2$ and $Q_2^2$ as long as $Q_3^2$ stays small. Using available input for the $\pi^0$-pole term, we find that the shift due to longitudinal SDCs on the isovector part of $a_\mu^\text{HLbL}$ is in the range \num[separate-uncertainty]{2.6\pm1.5e-11}. By including in the analysis also the isoscalar components, which the $\eta$- and $\eta'$-poles are assumed to dominate at low energies, we obtained that longitudinal SDCs increase $a_\mu^\text{HLbL}$ by \num[separate-uncertainty]{9.1\pm5.0e-11} in total. 
The quoted ranges encompass uncertainties in the low-energy input, perturbative corrections and fitting errors at asymptotic momenta, parametric variations of the functional form of the interpolants and of the matching surface, at which these functions are matched to the low-energy input, with the latter dominating the total uncertainty. Thus, according to our analysis, infinite towers of states heavier than \SI{1}{\GeV}, albeit crucial for the saturation of SDCs, give a relatively small contribution to $a_\mu^\text{HLbL}$ and this effect can be estimated with sufficient precision using our method. Conversely, states with masses around \SI{1}{\GeV} contributing significantly to the low-energy region play a decisive role also in a precision determination of short-distance effects.

Our result for the effects of longitudinal SDCs on $a_\mu^\text{HLbL}$ agrees with recent model estimates~\cite{BernSDCShort,BernSDCLong}, fulfills the accuracy goal set by the forthcoming experimental results and is significantly smaller than the earlier model result of Ref.~\cite{MV}, especially when up-to-date TFF input is used. Furthermore, neglecting issues concerning intrinsic model dependence and the fact that holographic QCD calculations in Refs.~\cite{HolographyVienna, HolographyItaly} do not completely saturate the SDCs, we find in agreement with these studies that the infinite tower of axials has a relatively small impact on the longitudinal part of $a_\mu^\text{HLbL}$ if the lightest multiplet is treated explicitly as a low-energy contribution.

It will be straightforward to incorporate in our approach model-independent information on further intermediate states as well as higher-order corrections to asymptotic expressions once these become available. Furthermore, our method can be generalized to the case of transversal SDCs. Therefore, it paves the way for a combination of all available low- and high-energy information on HLbL into one model-independent, accurate numerical estimate of this contribution to the muon $g-2$.

\begin{acknowledgements}
\vspace{0.2cm}

\noindent
We thank Gilberto Colangelo, Franziska Hagelstein, Martin Hoferichter, Laetitia Laub and Peter Stoffer for several useful discussions and comments on the manuscript. We also thank Nils Hermansson-Truedsson, Bai-Long Hoid, Simon Holz, Daniel Lechner, Josef Leutgeb, Pere Masjuan, Anton Rebhan and Pablo Sanchez-Puertas for helpful communications. J.L.\ is supported by the FWF-DACH Grant I 3845-N27 and by the FWF doctoral program Particles and Interactions, project no.\ W1252-N27.

\end{acknowledgements}

\appendix

\section{Convergence of the interpolants}
\label{App:Convergence}
In this appendix we discuss under which conditions our interpolants in Eqs.~(\ref{Eq:interpolant12}) and (\ref{Eq:interpolant3}) converge to the true $\PiPS$ as $N\to \infty$. To this end, we assume that $\PiPS$ is known exactly in the region below the matching surface, {\it i.e.}\ for $\Sigma < \Sigma^\text{match}(r,\phi)$, and that $\PiPS \to \PiPS[\text{asymp}]$ for asymptotic $\Sigma$.

The BTT scalar function $\PiPS$ is free of kinematic singularities and analytic except for poles and branch cuts for configurations where the real part of at least one $Q_i^2$ is negative. For fixed $(r,\phi)$ with $0 \le r < 1$ and $0 \le \phi < 2\pi$, $\PiPS$ is an analytic function of $\Sigma$ except for poles and branch cuts for $\Re(\Sigma) < 0$. $\PiPS[\text{asymp}]$ for fixed $(r,\phi)$ is also an analytic function except for isolated poles at $\Sigma \leq 0$ (see Eq.~(\ref{Eq:AsymPi})). The ratio $\PiPS / \PiPS[\text{asymp}]$ therefore has the same singularities as $\PiPS$ and has a pole at the zero of $\PiPS[\text{asymp}]$, which we assume to be at $\Sigma^\text{pole} < \Sigma^\text{match}$.
We can thus write the ratio as a Taylor series in $\Sigma^{-1}$ at $(\Sigma^\text{match})^{-1}$,
\begin{equation}
\frac{\PiPS(\Sigma)}{\PiPS[\text{asymp}](\Sigma)} = \sum_{i=0}^{\infty} a_i \left(\frac{1}{\Sigma} - \frac{1}{\Sigma^\text{match}}\right)^i\,.
\end{equation}
This series converges for $\Sigma^{-1} \in (2(\Sigma^\text{match})^{-1} - (\Sigma^\text{pole})^{-1},\allowbreak (\Sigma^\text{pole})^{-1})$ or equivalently for $\Sigma \in (\Sigma^\text{pole}, \infty)$ if the relation $\Sigma^\text{pole} < \Sigma^\text{match}/2$ holds, which will be checked below. Since $\PiPS / \PiPS[\text{asymp}] \to 1$ as $\Sigma \to \infty$, we also know that
\begin{equation}
\sum_{i=0}^\infty a_i \left(-\Sigma^\text{match}\right)^{-i} = 1\,.
\label{Eq:AppAsymp}
\end{equation}
We can thus write
\begin{equation}
\begin{aligned}
\PiPS(\Sigma) &= \PiPS[\text{asymp}](\Sigma) \sum_{i=0}^{\infty} a_i \left(\frac{1}{\Sigma} - \frac{1}{\Sigma^\text{match}}\right)^i \\
&= \PiPS[\text{asymp}](\Sigma) \sum_{i=0}^{\infty} b_i \Sigma^{-i}\,,
\end{aligned}
\label{Eq:AppAsympSeries}
\end{equation}
where the $b_i$ are linear combinations of the $a_i$ with coefficients depending on $\Sigma^\text{match}$. In particular, $b_0 = 1$
due to Eq.~(\ref{Eq:AppAsymp}). Eq.~(\ref{Eq:AppAsympSeries}) shows that $\PiPS[\text{int 1}]$ converges to the true $\PiPS$ for $N \to \infty$ if $\Sigma^\text{pole} < \Sigma^\text{match}/2$ for all $(r,\phi)$ in the HLbL integration domain and $\Sigma > \Sigma^\text{pole}$. In the applications of our method, $N$ is limited to rather low values since $\PiPS$ and its derivatives at the matching surface are determined only from the $\pi^0$, $\eta$, $\eta'$-poles (and additionally from the lightest axial in Sec.~\ref{Sec:NumAxials}).

Let us now examine under which conditions the zero in $\PiPS[\text{asymp}]$ occurs for $\Sigma^\text{pole} < \Sigma^\text{match}/2$. This relation is independent of the low-energy input but depends on the pseudoscalar mass in $\PiPS[\text{asymp}]$. In our reference interpolant we set $P(x, y) = 0$ in $\Sigma^\text{match}$ and $m^2 = \SI{0.5}{\GeV^2}$. For these choices the zero in $\Pipion[\text{asymp}]$ is at sufficiently low $\Sigma$ for all $(r, \phi)$ and $m^2$ can be reduced down to \SI{0.0019}{\GeV^2} without violating the requirement $\Sigma^\text{pole} < \Sigma^\text{match}/2$. For the iso-singlet cases there is no zero for positive $\Sigma$ in $\Pietaetap[\text{asymp}]$ allowing all values for $m$. This does not place a serious limitation on the values for $m$ we consider in the uncertainty estimation in Sec.~\ref{Sec:NumMatch}.

Since $\PiPS[\text{asymp}]$ for fixed $(r,\phi)$ in the integration domain has poles only for $\Sigma \le 0$ and $\PiPS$ has no kinematic zero, also the ratio $\PiPS[\text{asymp}] / \PiPS$ can be expanded in $\Sigma^{-1}$ at positive $(\Sigma^\text{match})^{-1}$. The same line of arguments thus proves the convergence of interpolant 2 in Eq.~(\ref{Eq:interpolant12}) for $N\to\infty$ and the requirement $\Sigma^\text{pole} < \Sigma^\text{match}/2$ is trivially fulfilled due to $\Sigma^\text{pole} = 0$. The convergence of interpolant 3 given in Eq.~(\ref{Eq:interpolant3}) also easily follows from that of interpolant 1, because the logarithmic term can be written as a Taylor series at finite $(\Sigma^\text{match})^{-1}$ so that the parameter $b_1$ in Eq.~(\ref{Eq:interpolant3}) is redundant as $N\to\infty$.

\bibliographystyle{spphys}
\bibliography{refs}

\end{document}